\documentclass[aps,prl,reprint,superscriptaddress,floatfix]{revtex4-2}
\usepackage{amssymb}
\usepackage{amsmath}
\usepackage{makecell}
\usepackage{color}
\usepackage[pdftex]{graphicx}
\usepackage{multirow}
\usepackage{float}
\usepackage{mathtools}
\usepackage{comment}
\usepackage{colortbl}
\usepackage{nicematrix}
\usepackage[toc,page]{appendix}

\usepackage{makecell}
\usepackage{enumerate}
\usepackage{mathtools}
\usepackage{stmaryrd}
\usepackage{enumitem}
\usepackage{textcomp}
\usepackage{gensymb}
\usepackage{float}

\def \equi#1{\mathrel{\mathop{\kern 0pt\sim}\limits_{#1}}} 

\usepackage{diagbox}

\usepackage[overload]{empheq}
\usepackage{cases}

\graphicspath{{FiguresPaper/}} 

\begin{document}

\title{Evolutionary Advantage of Diversity-Generating Retroelements in Switching Environments}

\author{L\'eo R\'egnier}
\address{Laboratory of Physics of the Ecole Normale Supérieure, PSL \& CNRS UMR8023, Sorbonne Universit\'e, Paris, France}
\author{Paul Rochette}
\address{Synthetic Biology, Institut Pasteur, Université Paris Cit\'e,  Paris, France.}
\author{Raphaël Laurenceau}
\address{Synthetic Biology, Institut Pasteur, Université Paris Cit\'e,  Paris, France.}
\author{David Bikard}
\address{Synthetic Biology, Institut Pasteur, Université Paris Cit\'e,  Paris, France.}
\author{Simona Cocco}
\address{Laboratory of Physics of the Ecole Normale Supérieure, PSL \& CNRS UMR8023, Sorbonne Universit\'e,  Paris, France}
\author{Rémi Monasson}
\address{Laboratory of Physics of the Ecole Normale Supérieure, PSL \& CNRS UMR8023, Sorbonne Universit\'e,  Paris, France}

\begin{abstract}
Diversity-Generating Retroelements (DGRs) create rapid, targeted variation within specific genomic regions in phages and bacteria. They operate through stochastic retro-transcription of a template region (TR) into a variable region (VR), which typically encodes ligand-binding proteins. Despite their prevalence, the evolutionary conditions that maintain such hypermutating systems remain unclear. Here we introduce a two-timescale framework separating fast VR diversification from slow TR evolution, allowing the dynamics of DGR-controlled loci to be analytically understood from the TR design point of view. We quantity the fitness gain provided by the diversification mechanism of DGR in the presence of environmental switching with respect to standard mutagenesis. Our framework accounts for observed patterns of DGR activity in human-gut \textit{Bacteroides} and clarifies when constitutive DGR activation is evolutionarily favored.
\end{abstract}

\maketitle

\begin{figure}[th!]
    \centering
    \includegraphics[width=0.9\columnwidth]{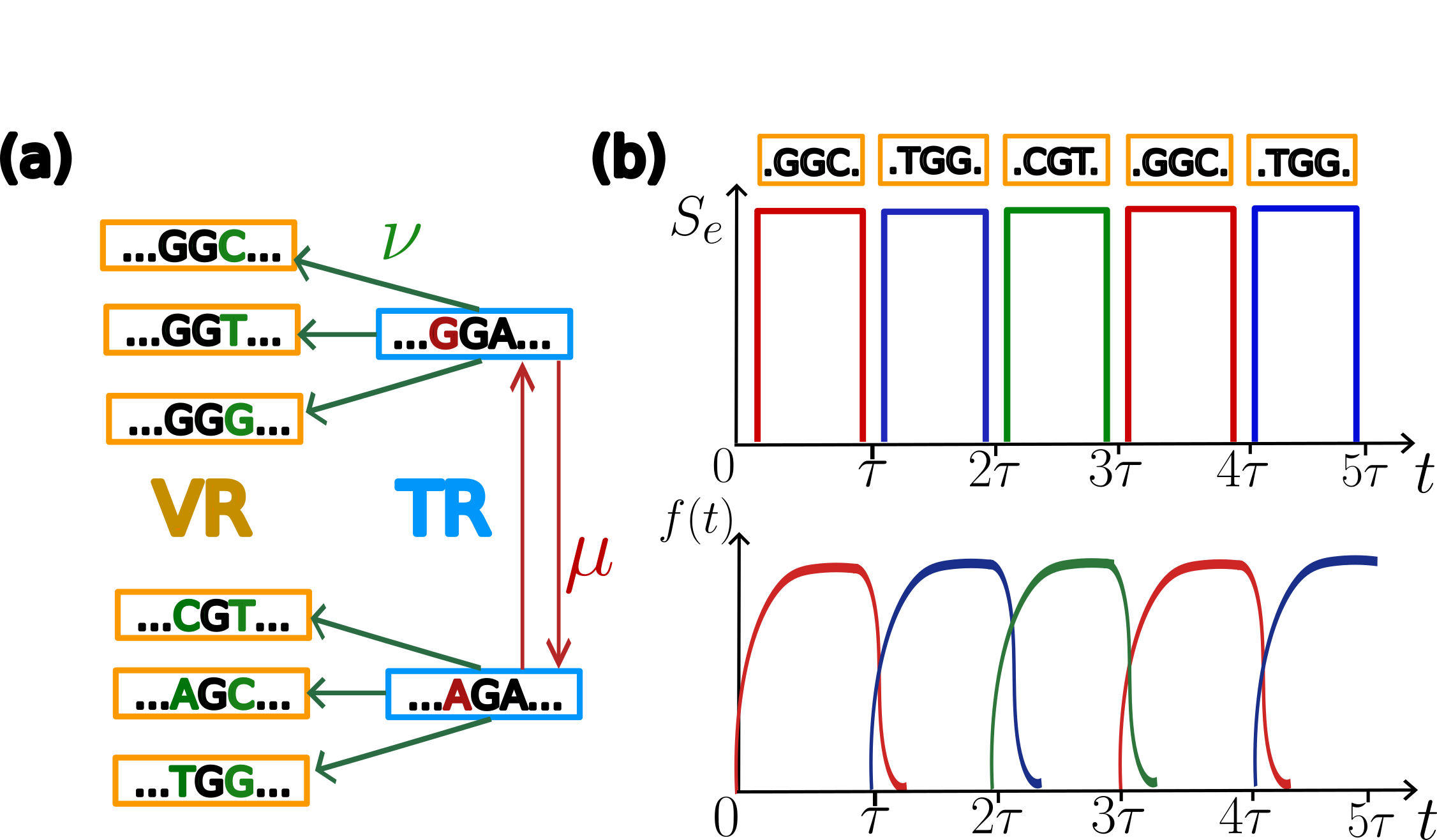}
    \caption{{\bf Illustration of the DGR system.} {\bf (a)} The main components of a DGR are shown. The template region (TR, blue) is copied into the variable region (VR, orange) at rate $\nu$, with mutations introduced specifically at adenine (A) positions. The TR  mutates at rate $\mu$. The fitness of an individual---defined by its (TR, VR) pair---depends solely on its VR sequence. 
{\bf (b)} Sketch of a fluctuating environment. Every interval $\tau$, the environment switches to favor a different target (VR) sequence. 
The population frequency $f(t)$ of the fittest sequence (with fitness $S_e$) then adapts to these shifts.}
    \label{fig:Intro}
\end{figure}

\noindent {\bf Introduction}.
Diversity-Generating Retroelements (DGRs), first identified in the \textit{Bordetella} phage \cite{Liu:2002}, represent an important class of hypermutation systems that enable rapid adaptation in phage and microbial populations. Recent work has clarified their structural organization \cite{Dai:2010,Handa:2025}, documented their diversity across microbial genomes \cite{Wu:2018,Roux:2021,Carrasco:2024}, elucidated their biological roles \cite{Vallota:2020,Carrasco:2024}, and highlighted their potential for engineering applications \cite{Laurenceau:2025,Unlu:2025}.  DGRs introduce mutations by copying a template region (TR) into a target variable region (VR) through an error-prone reverse transcription process \cite{Guo:2008,Guo:2011,Guo:2014} (Fig.~\ref{fig:Intro}{\bf (a)}). The mechanism is inherently asymmetric. While the TR slowly evolves through standard mutagenesis, the VR is quickly and repeatedly diversified and overwritten, with mutations concentrated at positions corresponding to a specific TR nucleotide (typically, Adenine). This localized hypermutating system can generate extensive phenotypic diversity at loci involved in host recognition, cell–cell interactions, or environmental sensing \cite{Vallota:2020,Carrasco:2024,Roux:2021}.  

Recent studies on the human gut microbiome \cite{Roux:2021,Gulyaeva:2024,Macadangdang:2025} illustrate the benefits of high DGR mutation rates \cite{Giraud:2001}. Infants are born with very few DGRs in their microbiome, but by one year, they reach adult-like levels. Remarkably, $\sim 72\%$ of the 388 parental DGRs were observed to switch their VR during this period. 
Some \textit{Bacteroides} species (notably \textit{B. ovatus} and \textit{B. finegoldii}) maintain high DGR activity: 
13--40\% of VRs diverge within 14 days. This suggests that diversification operates on a dedicated timescale far faster than standard mutation. 

While environmentally regulated DGR activation might explain infant microbiome diversification and context-dependent activity in many species \cite{Macadangdang:2025,Macadangdang:2022,Roux:2021}, it does not address a more fundamental question: under what conditions is a DGR-based hypermutating system actually advantageous and thus maintained rather than lost? Because DGR activity depends on a functional reverse transcriptase, mutations that impair this enzyme could in principle silence diversification, highlighting a potential evolutionary fragility of such systems. The long-term selective benefit of such constitutively active diversification is not immediately obvious and requires explicit mechanistic modeling. Although hypermutation has been studied in other biological contexts \cite{Martincorena:2013,Lynch:2016,Ram:2012}, the evolutionary forces governing the dynamics in systems with the distinctive architecture of DGRs remain largely unexplored.

\noindent
{\bf DGR model and evolutionary dynamics.} Hereafter, a DGR is defined by the nucleotidic sequences  of its {\bf VR} and {\bf TR}, of equal length $L$ (Fig.~\ref{fig:Intro}{\bf (a)}). In the following, we use bold symbols when refering to the precise sequences of nucleotides. Under standard point mutation process, each nucleotide in the {\bf VR} and {\bf TR} may mutate with rate (probability per unit of time) $\mu$. In addition, the {\bf VR} varies due to the DGR replacement and hypermutation mechanism, which happens at rate $\nu \gg \mu$. In practice, we assume that a new {\bf VR} replaces the current {\bf VR} after being randomly drawn from the {\bf TR}. In this random process, nucleotides $C,G,T$ in the {\bf TR} are copied as such in the {\bf VR} at the same positions, while $A$'s are equally likely to mutate into $C,G,T$ or to remain unchanged \footnote{We do not need to consider partial correction of the {\bf VR} by the repair protein MutS. As observed in Ref.~\cite{Macadangdang:2025} the DGR diversification mechanism is all-or-nothing: the {\bf VR} is either entirely corrected into the initial {\bf VR}, or entirely kept as it is.}. 

As proteins are expressed from the {\bf VR}, the fitness (growth rate) $S({\bf VR},t)$ of the full DGR depends exclusively on the nucleotide sequence of the {\bf VR}. For simplicity, we assume that $S$ is additive across sites (no epistasis) \cite{Halpern:1998,Wylie:2011}, and that fitness contributions depend solely on the nucleotide identity at each position rather than the encoded amino acids. The fitness at time $t$ is given by
\begin{align}\label{eq1}
S(\text{\bf VR},t) = \sum_{i=1}^L s_{i, \text{VR}_i}(t) \ ,
\end{align}
where $s_{i,q}$ denotes the  contribution  of a nucleotide $q=A,T,C,G$  at position $i$ along the {\bf VR} sequence. To model  environmental changes, we assume below that the $s$ factors vary with time $t$ on a characteristic switching time $\tau$, as represented in Fig.~\ref{fig:Intro}{\bf (b)}.

A key insight is that selection acts on the DGR on {\it two distinct timescales}:
\begin{itemize}[leftmargin=5.5mm]
    \item \textbf{Fast (at the VR level):} {\bf VR} sequences diversify through repeated overwriting by the TR at rate $\nu$ and compete under environmental fluctuations.  The pattern of Adenines (A) on the {\bf TR} determines which {\bf VR} genotypes are accessible.  
    \item \textbf{Slow (at the TR level):} the {\bf TR} sequence mutates  at the much lower per-base rate $\mu$. {\bf TR} variants are indirectly selected \emph{through} the {\bf VR}'s they generate: the mean fitness of the {\bf VR} population shapes the long-term evolutionary potential of the {\bf TR}. 
\end{itemize}

This separation of timescales is strongly supported by experimental observations (see below) and allows for an analytical treatment of the DGR dynamics.
In the following, we first analyze \textbf{VR selection}, showing how the effective growth rate depends on the DGR replacement rate $\nu$ relative to environmental switching $\tau$. Second, we address \textbf{TR selection}, explaining how selection acts on the TR sequence itself, and in particular, we derive the conditions under which the diversifying nucleotide (Adenine) is evolutionarily stable. In this framework, loss of DGR activity is modeled exclusively through selection acting on the TR sequence; we do not consider deactivation arising from mutations in the reverse transcriptase. Lastly, we explore the effects of the TR length on the selection dynamics, identifying the regimes in which DGR-mediated diversification is most advantageous. This framework explains experimental observations and identifies conditions under which DGRs are maintained, regulated, or lost in natural systems. \\

\noindent
{\bf Experimental background.} The authors of~\cite{Macadangdang:2025} studied five Bacteroides DGR, and found two species that maintained a high level of DGR activity both {\em in vitro} and {\em in vivo} in the absence of selective pressure (as evidenced by sustained activity in monoconolized mice). After $t=$14 days, a fraction $\varphi$ of VR ranging from 13 to 40 \% had diverged from the initial ones. 

Within our modeling framework, this fraction is simply given by $\varphi=1-e^{-\nu t}$ due to the lack of selection ($s$ independent of the nucleotide content), see details in Supplementary Materials (SM) \footnote{See Supplemental Material at [URL will be inserted by publisher] for analytical details and additional figures.} Sec.~1. Matching this formula with the experimentally observed fraction, we infer the diversification rates for these two Bacteroides  to be between $\nu=10^{-2}~day^{-1}$ and $\nu=4 \times10^{-2}~day^{-1}$, see Table~\ref{tab}.

\begin{table}[h!]
\centering
\begin{tabular}{lcc}
\hline
\textbf{Quantity} & \textbf{Notation} & \textbf{Range (day$^{-1}$)} \\
\hline
DGR diversification rate & $\nu$ & $10^{-2}$ -- $4\times 10^{-2}$ \\
[-2mm]
Growth rate & $S_e$ & $1$ -- $10$ \\
[-2mm]
Growth rate per variable site & $s$ & $10^{-1}$ -- $10$ \\[-4.5mm] 
\multirow{2}{*}{Spontaneous mutation rate} & \multirow{2}{*}{\centering$\mu$} & \multirow{2}{*}{\centering$10^{-9}$ -- $10^{-8}$} \\
[-0.5mm]
\multicolumn{1}{c}{per base} & & \\
[-3mm]
Mutation rate for full VR & $\mu L$ & $10^{-7}$ -- $10^{-5}$ \\
\hline
\end{tabular}
\caption{Approximate rates relevant for DGR dynamics in \textit{Bacteroides}.}
\label{tab}
\end{table}

As a point of comparison, the actual doubling time of Bacteroides is around $2$ hours in optimal conditions and up to a day \cite{Frantz:1979,Zhao:2019} (in situations where selection plays a role), resulting in a growth rate $S_e$ between $1$ and $10~ day^{-1}$. Because the spontaneous mutation rate per generation per base is about $10^{-9}$, the rate $\mu$ of spontaneous mutation per base is of the order of  $10^{-9}-10^{-8}~ day^{-1}$ \cite{Zhao:2019,Garud:2019}. Consequently, the rate $\mu L$ of spontaneous mutation apparition (without DGR) in the region where selection occurs (the VR encoded part, approximately 100-1000 bases) ranges  between $10^{-7}$ and $10^{-5}~ day^{-1}$. The effective number of sites 
$L$ should be restricted to codons containing adenines within the hypermutation interval, not the entire sequence, as the phenotype is defined by the protein product encoded by the VR. This effective number $L$ of variable sites can be estimated from the data of Ref.~\cite{Roux:2021} to be between $1$--$8$ (interval which contains $95\%$ of the data, see SM Sec.~2).  Consequently, there is a huge time scale separation (at least 3 orders of magnitude) between the estimated DGR rate, the replication rate and the rate at which spontaneous mutations appear in a sequence: $\mu L \ll \nu \leq  S_e$. This is a strong indication that DGR mutation and selection work on a time scale different from the classical selection mechanism. 
\\

\noindent
{\bf Fast dynamics of the {\bf VR} population at fixed {\bf TR}.} On time scales $< 1/(\mu L)$, the TR is effectively fixed, while the VR evolves under selection and is periodically overwritten via DGR-mediated diversification at rate $\nu$. Because the model assumes independent sites, we adopt a per-base perspective: we first focus on a single site ($L=1$), and then generalize the results to $L > 1$.

Nucleotides $A,C,G,T$ are labeled as $1,2,3,4$ ($Q=4$ possible states).
For the sake of simplicity, we consider a cyclic environmental dynamics of period $Q\tau$. Here $\tau$ is the switching time. For $t\in [q'\tau, (q'+1)\tau]$ with $q'=1, ..., Q$, the preferred nucleotide is $q^*(t)=q'$ and confers a fitness advantage $S(q',t)=s>0$ (as $L=1$) with respect to the other $Q-1$ ones, {\em i.e.} $s_{1,q}(t)=s\,\delta_{q,q^*(t)}$ in Eq.~(\ref{eq1}). This model corresponds to a classical framework of phenotypic switching \cite{DeGroot:2023,Kussell:2005,Taitelbaum:2020}. 

\begin{figure}[th!]
    \centering
    \includegraphics[width=\columnwidth]{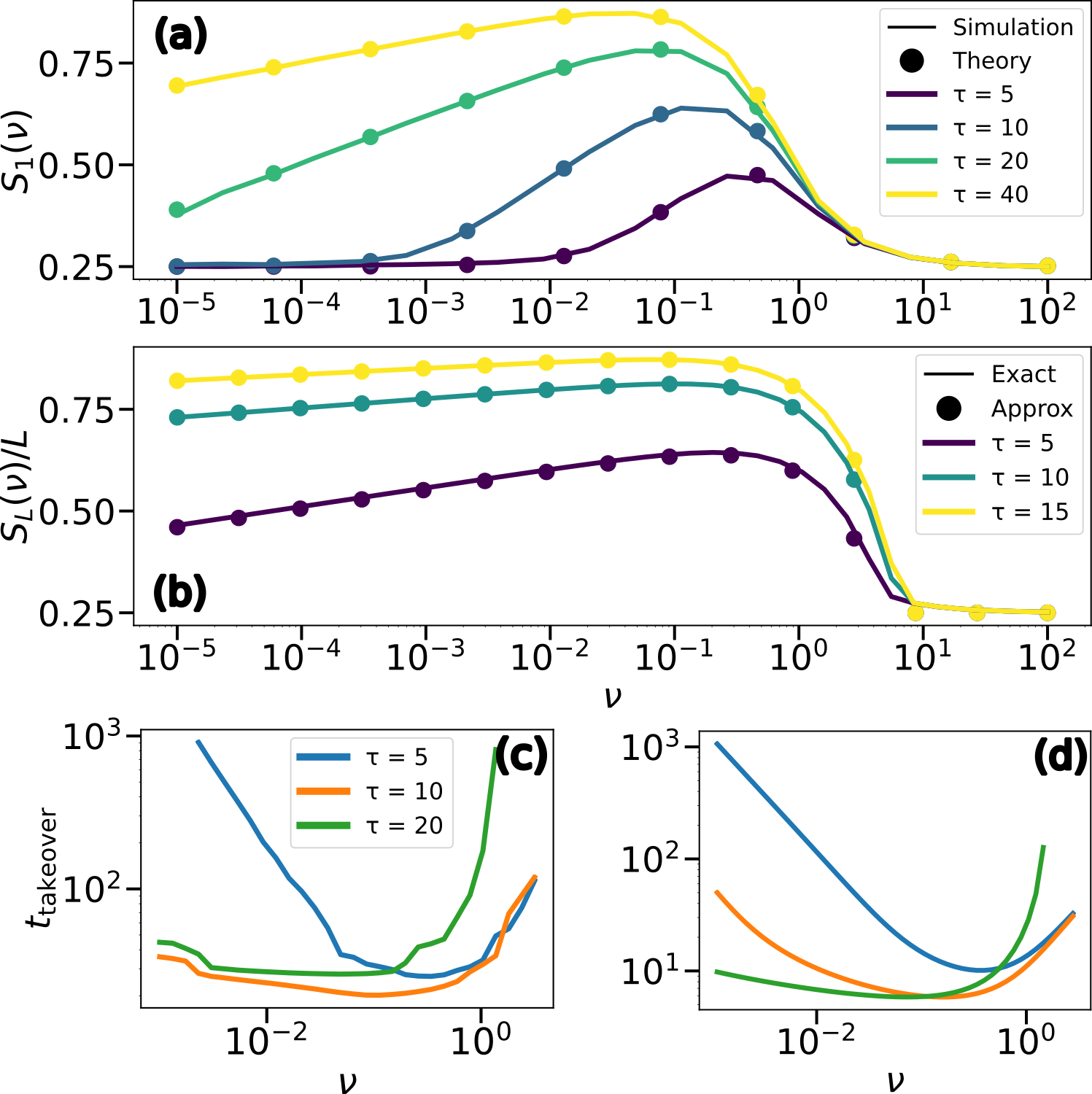}
    \caption{{\bf VR selection.}{\bf (a)} Effective growth rate as a function of the switching rate $\nu$ for $L=1$. {\bf (b)} Same as (a) for  length $L=10$.  Takeover times for the DGR system   estimated from simulations (more than 99.5\% of the population) {\bf (c)} and from Eq. \eqref{takeover} {\bf (d)} vs. diversification rate $\nu$ for $L=1$. Parameter values: $Q=4$, $s=1$,  population size $=10^{8}$, $\mu=10^{-5}$.}
    \label{fig:VR_selection}
\end{figure}

The population sizes $n_q(t)$ of ${\bf VR}$ carrying nucleotides $q$ at time $t$ obey the Wright-Fisher coupled equations:  
\begin{align} \label{eqdyn}
    \frac{{\rm d}n_q(t)}{{\rm d}t}&=\left(s\, \delta_{q,q^*(t)}-\nu\right)n_q(t)+\frac{\nu}{Q}\sum_{p=1}^Q n_{p}(t) \ .
\end{align}
Standard mutagenesis (no DGR) is described by the same equations with $\nu$ replaced with $\mu$. This set of linear differential equations can be solved as explained in SM  Sec.~3. Formally speaking, after each cycle of duration $Q\tau$, the $Q$-dimensional vector of population sizes, $\boldsymbol{n}=\{n_q\}$, is multiplied by a $(Q\times Q)$-dimensional matrix $\boldsymbol{M}$, whose entries encode the transition rates appearing in Eq.~\eqref{eqdyn}, see details in SM. Due to the periodicity of the time evolution, the effective growth rate is given by $S_1=\ln \Lambda/(Q \tau)$, 
where $\Lambda$ is the largest eigenvalue of $\boldsymbol{M}$.
This growth rate is plotted in Fig.~\ref{fig:VR_selection}{\bf (a)} as a function of the mutational rate and is in excellent agreement with the results of numerical simulations for finite population size.

As observed from Fig.~\ref{fig:VR_selection}{\bf (a)}, the growth rate $S_1$ is maximal when the rate $\nu$ is roughly equal to $1/\tau$ \cite{DeGroot:2023,Kussell:2005}, and largely exceeds its asymptotic value $s/Q$ at both large and small $\nu$. In each of these limits, a fixed (small $\nu$) or a random (large $\nu$) nucleotide is best for a fraction $1/Q$ of the time, explaining this expression. Notice that, for large $\tau$, $S_1$ decays slowly as the mutation rate $\nu$ is lowered below its optimal value $\sim 1/\tau$: the system has time to adapt to the slowly fluctuating environment even though mutations become rarer. 

The above results remain qualitatively correct for $L>1$ variable sites, {\em i.e.} for a {\bf TR} with $L$ Adenines.  We assume that two successive environments select for sequences with different states at every position. Without DGR, spontaneous mutations happen at rate $\mu$ independently on each site, and the effective fitness is just $L S_1(\mu)$. With the DGR mechanism, all {\bf VR} bases are simultaneously redrawn at each mutational event, introducing correlations between the sites. The effective fitness with  diversification rate $\nu$ reads
\begin{align}
\label{eq:Fitness_L>1}
    S_L(\nu) \approx \max \left(\frac{s L}{Q}, s L-\nu+\frac \nu{Q^{L}}+\frac{\ln \left(\frac{2\nu}{sL Q^L} \right)}{\tau} \right) \; ,
\end{align}
see SM Sec.~3 for details about the derivation. Results are shown in Fig.~\ref{fig:VR_selection}{\bf (b)}, and are qualitatively similar to the $L=1$ case. 

Our analysis unveils the condition under which a population of organisms equipped with DGR ultimately grows much faster than one lacking it (see also phase diagrams in SM Sec.~3).
In particular, when $\nu \sim 1/\tau$, there is a consistent advantage to the DGR mechanism regardless of the value of $S_L$. However, if $S_L$ is too small, the takeover time of the DGR-equipped population,
\begin{align}\label{takeover}
t_\text{takeover} \approx \frac{1}{S_L(\nu) - L\, S_1(\mu)} \ 
\end{align}
may become prohibitively long. Two conditions must be met to avoid large $t_\text{takeover}$ values, as illustrated in Fig.~\ref{fig:VR_selection}{\bf (c)} and {\bf (d)} for $L=1$. First, $\nu$ should not be larger than $s$, as replacement by DGR-generated variants would occur so frequently that the system would gain little advantage. Second, $\tau $ must not be exceedingly larger than $1/ s$, as low spontaneous mutation rate $\mu$ would be sufficient to track very slow environmental changes, as mentioned above. 
These conditions for the success of DGR appear to be met in experimental situations \cite{Macadangdang:2025}, where we infer $\nu  \lesssim S_e$. This condition stays true considering $L$ below 8 as the effective number of variable sites and $s=S_e/L$. (Table~\ref{tab}), ensuring both sufficient positive selection and high growth rate.\\

\noindent
{\bf Slow dynamics of the TR.}  We now allow the nucleotides in the {\bf TR} to mutate at a rate $\mu$. A detailed treatment of the Wright-Fisher equations describing the evolution of the population of $\bf (TR,VR)$ pairs in a fluctuating environment is provided in SM Sec.~4.  We hereafter summarize the main results and provide intuition that supports them.

\begin{figure}[th!]
    \centering
    \includegraphics[width=\columnwidth]{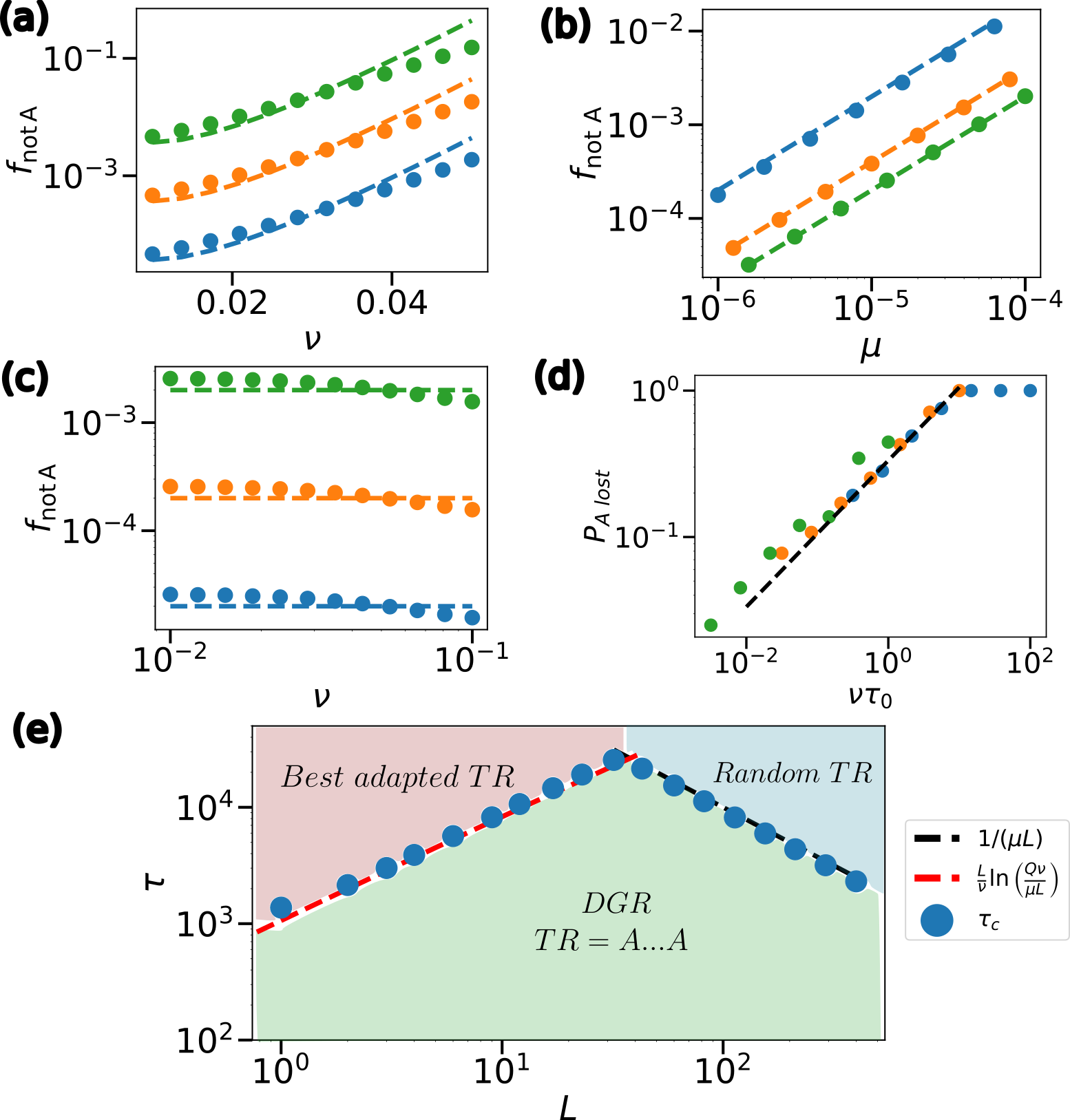}
    \caption{{\bf TR selection.} {\bf (a)} to {\bf (c)}. Fraction $f_{not~A}$ of sequences in the $\bf TR$ population carrying non-adenine nculeotides.  {\bf (a)} $\nu\tau\gg 1$ ($\tau=200$,  $\mu=10^{-7}$, $10^{-6}$, and $10^{-5}$); {\bf (b)} $\nu\tau= 1$ ($\nu=1/\tau=10^{-2}$, $5\cdot 10^{-2}$ and $10^{-1}$);  {\bf (c)} $\nu\tau\ll 1$ ($\mu=10^{-6}$, $10^{-5}$ and $10^{-4}$ and $\tau=5$). Increasing values of  $\nu$ or $\mu$ are shown in blue, orange and green successively. Dashed lines correspond to the expressions of Eq.~\eqref{eq:exp_nu_tau} in the respective regimes.    {\bf (d)} Probability $P_{A~lost}$ that the fraction of $\bf TR$ with an $A$ (initially, equal to 1) is lower than $1/Q=1/4$ as a function of $\nu \tau_0$ where $\tau_0$ is the typical duration of an environment. Dashed line corresponds to $\left( \nu \tau_0 \right)^\beta$. We take $\mu=10^{-4}$, $\beta=1/2$ and $\nu=10^{-3}$, $10^{-2}$ and $10^{-1}$. {\bf (e)} Phase diagram in the $(L,\tau)$ plane. Dashed lines show the theoretical prediction for $\tau_c$ in Eq.~\eqref{eq:tau_c} Blue points are determined by solving the dynamics for infinite population size. Parameters are $\mu=10^{-6}$ and $\nu=10^{-2}$.}
    \label{fig:TR_selection}
\end{figure}

Informally speaking, a {\bf TR} sequence can be evolutionarily fit in two distinct ways.
First, it may carry many $A$'s that allow the $\bf VR$ to diversify and adapt to any environment. However,  continual diversification of the $\bf VR$ at a rate $\nu$  leads to the frequent loss of adaptation in stable contexts. Second, the {\bf TR} may carry non-Adenine nucleotides that are well-adapted to the current environment; their copies in the {\bf VR} then provide a fitness advantage and are far less mutable (rate $\mu \ll \nu$). 
Analysis of the evolutionary dynamics of a population of sequences of length $L=1$ allows us to assess how these two factors may concur to push the $\bf TR$ to eventually loose its Adenines \footnote{Notice that if the environment favors $A$ on the $\bf VR$, then this adapted nucleotide can be produced only by a $A$ on the $\bf TR$. This case does therefore not contribute to the loss of Adenines on the $\bf TR$. Further details can be found in SM Sec.~3.}. In particular, we find that the overwhelming majority of $\bf TR$ sequences either carry a $A$, or do not carry a $A$ but are adapted to the current environment. The fraction of the latter is approximately given by
\begin{align}\label{eq:exp_nu_tau}
    f_\text{not A}=\frac{\tilde \mu}{\tilde \nu^2 \, \tau} \left(e^{\tilde \nu \tau}-1-\tilde \nu \tau\right)\ ,
\end{align}
where $\tilde \mu = Q\, \mu$ and $\tilde \nu = (1-\frac 1Q)\, \nu$. 
We deduce from the previous formula the existence of three regimes depending on the value of $\tilde \nu$ and $\tau$: $ f_\text{not A}\sim  \tilde \mu\,\exp\left({\tilde \nu \tau}\right) / (\tilde \nu^2\tau)$ for $\nu \gg 1/\tau$, $\tilde \mu/\tilde \nu $ for $\nu \sim 1/\tau$, $\tilde \mu\, \tau$ for $\tilde \nu \ll 1/\tau$.
We compare this approximation  with simulations in Fig.~\ref{fig:TR_selection}\textbf{(a)}, {\bf (b)}, and {\bf (c)}, confirming its validity. Imposing $f_\text{not A} \ll 1$ gives the conditions that allow the DGR mechanism to be favorable with respect to standard spontaneous mutational processes. If the switching time $\tau$ of the environment is tightly regulated, that is, if the changes in the environment are periodic, the DGR mechanism gives a fitness advantage provided that
\begin{equation}
    \tilde \mu \ll \tilde \nu \sim \frac{1}{\tau} \ .
\end{equation}

However, the condition above is not sufficient in the case of an environment with a wide distribution of switching times. Adenines in the $\bf TR$ become strongly disfavored if $\tau$ may significantly  exceed $1/\tilde \nu$, see Eq. \eqref{eq:exp_nu_tau}. For example, consider the case of $\tau$ algebraically distributed, with typical time-scale $\tau_0$ and exponent $\beta$, {\em i.e.} with cumultative probability $\mathbb{P}(\tau\geq t)=\left(\tau_0/t \right)^\beta$. The adenine is lost with a probability of the order of $\left( \tilde \nu \tau_0\right)^\beta$, condition which corresponds to $f_\text{not A}\sim 1$. This theoretical scaling is confirmed in Fig.~\ref{fig:TR_selection}{\bf (e)} by simulations estimating the probability $P_{A~lost}$ that, by the end of the environmental bout of random duration $\tau$, the population of $\bf TR$ carrying $A$ no longer dominates (fraction smaller than $1/2$).  This example shows that strongly (algebraically) variable environmental duration can be disastrous for the maintenance of the DGR mechanism, despite the fact that the typical timescale $\tau_0$ of these environmental duration still obey $\tilde \nu \tau_0 \approx 1$.\\

The above analysis can be extended to $L>1$. To do so, we examine the conditions that lead to the loss of sequences with all $L$ sites occupied by adenines. As seen previously, this maximally-diversifying $\bf TR$ competes with the best adapted sequence in the current environment, whose population grows as  $\left( \frac{\mu L}{Q\nu}\right)^Le^{\nu t}$: the prefactor corresponds to the $L$ spontaneous mutation events needed to reach the best adapted $\bf TR$, and the exponential term the gain in losing the DGR (SM Sec.~4). In addition, random sequences invade the population, with a fraction growing as $e^{\mu L t}$. As a consequence, the all-$L$ DGR is outcompeted when the time $\tau$ exceeds 
\begin{align} \label{eq:tau_c}
    \tau_c(L) &=\min\left[ \frac{L}{\nu}\, \ln\left(\frac{\nu Q}{\mu L}\right) ,\frac{1}{\mu L} \right]\ . 
\end{align} 
see Fig.~\ref{fig:TR_selection}{\bf (e)}. The phase diagram in the $(L,\tau)$ plane shows the existence of three regimes, corresponding to the three types of dominant $\bf TR$ sequences. 
The DGR phase is best stable over time for the optimal length
\begin{align}
    L^*\sim \sqrt{\nu/\mu} \; ,
\end{align}
up to a logarithmic factor.
This result stresses that having extended variable regions in the $\bf TR$ can be beneficial for the resilience of the DGR system, but that too long $\bf TR$ are not robust against spontaneous mutations. This is consistent with the relatively low number $L$ (up to $10^2$) of variable sites found in \cite{Roux:2021,Macadangdang:2025} and displayed in SM Sec.~2. We stress that our analysis focuses on the potential loss of DGR due to the TR dynamics only. Other mechanisms, such as reverse transcriptase inactivation, are left for future studies.\\

\noindent
\textbf{Conclusion.} Our analysis unveils to what extent and conditions the DGR system offers fitness advantages over standard spontaneous mutation processes in frequently changing environments. If the environmental context is stable or does not vary quickly enough, the DGR becomes progressively inoperant through the loss of the Adenines in the {\bf TR} after some critical time we have estimated. Our study therefore emphasizes the need for being able to tune the activation and desactivation of the DGR mechanism \cite{Nimkulrat:2016,Macadangdang:2025}, despite its inherent cost {\em e.g.} due to the sensing of the environment fluctuations. Organisms can  enable/disable the DGR apparatus in multiple ways: some might down-regulate diversification when it is unnecessary \cite{Macadangdang:2025,Roux:2021}, others may effectively silence the system through mutations that impair the reverse transcriptase \cite{Guo:2011,Naorem:2017,Macadangdang:2025}, while engineered systems can bypass the VR entirely and reintegrate cDNA directly into the TR \cite{Laurenceau:2025}. An additional relevant context include coevolving DGR populations \cite{Buckling:2002} in which the environmental switching time $\tau$ is replaced by the inverse DGR switching rate $1/\nu$ of the other population. We could also consider scenarios in which finite population size modulates the benefits of hypermutation \cite{Lynch:2016}. All of these possibilities constitute promising directions for future modeling efforts.

Future analysis should take into account the existence of  mutational biases inherent to DGRs and their consequences for protein-level selection. Because selection acts on amino acids, the position of adenines within codons can substantially affect the number of accessible amino-acid variants (for example, AAC codons can diversify into 15 amino acids, whereas having an adenine only at the third position yields negligible variation). These positional effects, together with biases in DGR A-to-N substitutions \cite{Laurenceau:2025,Macadangdang:2025,Unlu:2025}, shape the landscape of possible receptor or ligand variants. Exploring how diversification proceeds under these alternative architectures and mutational regimes would be interesting.

\bibliography{biblio}
\bibliographystyle{apsrev4-1}
\end{document}


\RestyleAlgo{ruled}
\title{SUPPLEMENTAL MATERIAL\\
A model of Diversity Generating Retroelements population dynamics.}

\maketitle

\tableofcontents

\section{Experimental background}
In this section, we explain the calculations of the {\bf Experimental background} section of the main text. \\
The setup is the following: the experiment starts with a population $n_0(0)$ of individuals with the same Variable Region (VR). At a rate $\nu$, the Template Region (TR) go through the Diversity Generating Retroelement (DGR) mechanism which leads to the replacement of the initial VR by a new one. We note the population in this category $n_1(t)$, population initially empty.\\
Because there is no selection against the VR in the experiment of Ref.~\cite{Macadangdang:2025}, we suppose that both populations grow at a same rate $S_e$. Finally, because it is unlikely for the system to produce the same VR as the initial one through the DGR mechanism, the number of possible VR growing exponentially with the number of Adenines, we consider that an individual in population $1$ stays there. Of note, even for small number of Adenines, this changes the estimated rate $\nu$ by a prefactor $1-1/Q$ at most (probability to produce the same nucleotide with only one diversifying nucleotide). \\
From this, we have the differential equations describing the time evolution of the system: 
\begin{align}
    \dot{n}_0(t)=(S_e-\nu) n_0(t)\\
    \dot{n}_1(t)=S_e n_1(t)+\nu n_0(t) \; .
\end{align}
The solution of this system of equations is given by: 
\begin{align}
   n_0(t)=n_0(0)e^{(S_e-\nu)t} \; ,
   n_1(t)=n_0(0)\left(e^{S_e t }-e^{(S_e-\nu)t}\right) \; ,
\end{align}
from which we deduce the expression of the fraction $\varphi$ of the population with distinct {\bf VR} from the original one as a function of time given in the main text:
\begin{align}
    \varphi =\frac{n_1(t)}{n_0(t)+n_1(t)}=1-e^{-\nu t} \; .
\end{align}

\section{Estimation of the number $L$ of variable sites}
\begin{figure}[th!]
    \centering
    \includegraphics[width=0.5\linewidth]{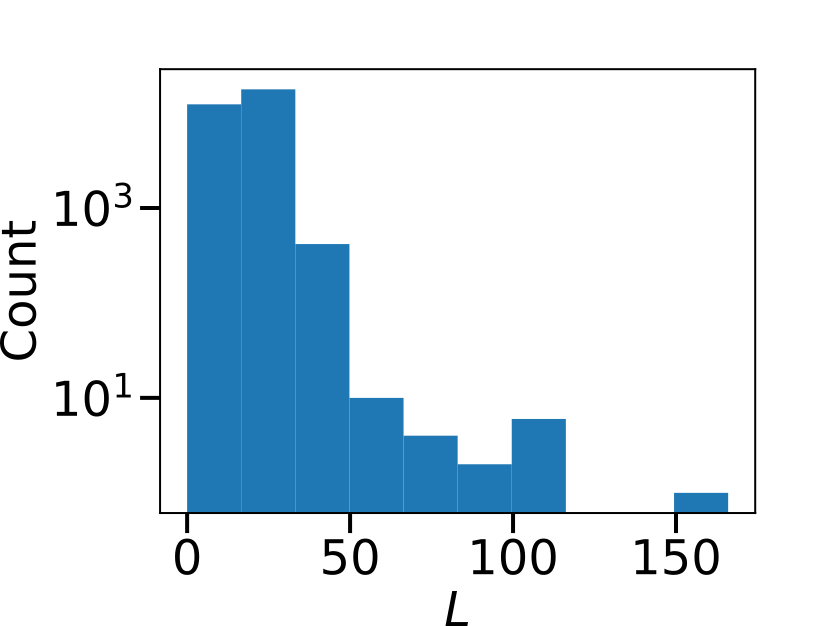}
    \caption{{\bf Number of variable sites distribution.} Histogram of the number $L$ of codons containing an Adenine in the region detected as DGR from Ref.~\cite{Roux:2021}, with the additional constraint that the codon is in an interval where DGR mutations have been detected.}
    \label{fig:Count_L}
\end{figure}

We do an estimate of the number of variable sites in the region undergoing DGR via the dataset assembled in Ref.~\cite{Roux:2021}. Because it is not the whole TR that will go through the hypermutation mechanism of the DGR, we consider only codons which are between two mutations (detected in the VR). All codons having an Adenine are then considered to be potential variable sites (even though there was no mutation in the VR at those positions). The histogram of the count $L$ of such variable sites is presented in \ref{fig:Count_L}. We obtain that $95\%$ of sequences have less than $L=8$ variable site in the region undergoing the DGR hypermutation mechanism.

\newpage

\section{Fast dynamics of the VR population at fixed TR}
\subsection{$L=1$}
As detailed in the main text, the population vector $\boldsymbol{n}=\left(n_q \right)_{0\leq q\leq Q-1}$ of the $Q$ nucleotides obeys the matrix equation
\begin{align}
    \frac{{\rm d}\boldsymbol{n}}{{\rm d}t}=A_{q^*(t)}  \boldsymbol{n} \; .
\end{align}
The $A_{q^*(t)}$ matrix corresponds to $\left(A_{q^*(t)}\right)_{q_1,q_2}=\delta_{q_1,q_2}(s \delta_{q_1,q^{*}(t)}-\nu)+\nu/Q$: all states mutate equally into any other at rate $\nu$, and only the nucleotide $q^*(t)$ (which changes every time $\tau$) has a growth rate of $s$. \\
The solution is formally expressed as $\boldsymbol{n}(Q\tau)=\prod_{q=0}^{Q-1}\exp \left(\tau A_{q}  \right)\boldsymbol{n}(0)=M\boldsymbol{n}(0)$. At large times $T=k Q\tau$,  the population has been through $k$ growth cycles, and the population has been enriched by a factor $M^k$. To get the growth rate of the population under these conditions, we find the largest eigenvalue $\Lambda$ of $M$, which gives the effective population growth rate $S_1(\nu)$ such that
\begin{align}
    e^{S_1(\nu) T}=e^{S_1(\nu) k \tau Q}\approx \sum_p (M^k \cdot \boldsymbol{n}(0))_p \propto  \Lambda^k 
\end{align}
leading to
\begin{align}
    S_1=\frac{\ln \Lambda }{Q \tau}
\end{align}
as explained in the main text.

\begin{figure}[th!]
    \centering
    \includegraphics[width=0.8\columnwidth]{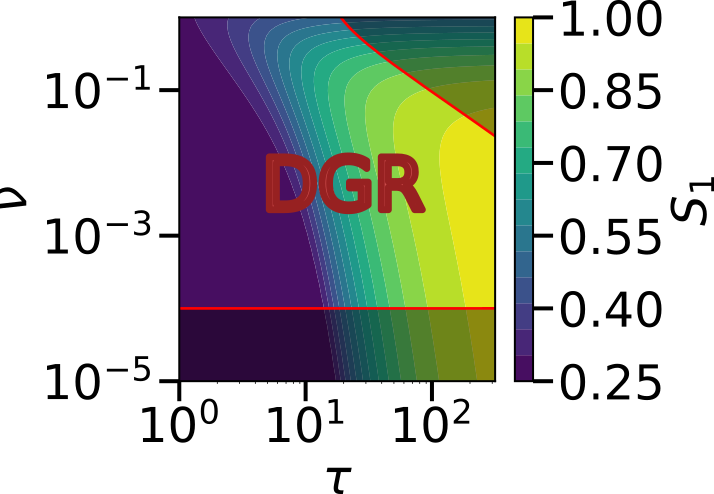}
    \caption{Phase diagram of the DGR vs non-DGR system as a function of $\nu$ and $\tau$, for $\mu=10^{-4}$. The region where  $S_1(\nu)>S_1(\mu)$ is delimited by the red contour. The region $S_1(\nu)<S_1(\mu)$ is darkened.}
    \label{fig:PD_tau}
\end{figure}

We show in \ref{fig:PD_tau} the phase diagram in $(\tau,\nu)$ which delimits the parameter space region where the DGR system dominates the non DGR one {\it i.e.} $S_1(\nu)>S_1(\mu)$.

\subsection{$L>1$ }
\subsubsection{$Q=2$: perturbation theory}
In this section we present the method for deriving the large $L$ effective fitness of a DGR system of length $L$, replacement rate $\nu$ and $Q=2$ nucleotides (let's note them $0$ and $1$). \\
During the time interval $[0, \tau]$, the system selects for $0$s: the fitness/growth rate of a sequence with $k$ zeros is simply $s k$. Similarly, during the time interval $[ \tau, 2\tau]$, the system selects for $1$s: the fitness/growth rate of a sequence with $k$ 0s (and $L-k$ 1s) is simply $s(L-k)$. Then due to the DGR system, each sequence is replaced at rate $\nu$ by a new one, uniformly distributed among any of the $2^L$ possible ones. We fix the scale of time with $s=1$. The system is cyclic of period $2\tau$.\\
We are interested in the effective growth rate of the system: if we note $A_0$ (resp. $A_1$) the transition matrix during the first time interval (resp. the second one), then the evolution of the population vector $\boldsymbol{n}$ of size $2^L$ indexed by the sequence (e.g. $11000$ for $L=5$) is given by
\begin{align}
    \boldsymbol{n}(2\tau)=\exp\left(\tau A_1\right) \cdot \exp\left(\tau A_0\right) \cdot \boldsymbol{n}(0) \; .
\end{align}
Thus, if we note $\Lambda_L$ the dominant eigenvalue of $\exp\left(\tau A_1\right) \cdot \exp\left(\tau A_0\right)$, the effective growth rate $S_L(\nu)$ of the population is given by:
\begin{align}
    S_L(\nu)=\frac{\ln \Lambda_L}{2 \tau} \; .
\end{align}
The objective is to compute $S_L(\nu)$. 
\paragraph{Reduction of the matrix size/Removal of degeneracy.}
At first, let's consider the exact form of $A_0$ and $A_1$. They are both $2^L$ square matrices, 
\begin{align}
    A_0&= \text{diag}\left(s \text{(\# of 0) }-\nu  \right) + \frac{\nu}{2^L}\boldsymbol{1} \cdot \boldsymbol{1}^\dagger=D_0+ \frac{\nu}{2^L}\boldsymbol{1} \cdot \boldsymbol{1}^\dagger \\
    A_1&=\text{diag}\left(s \text{(\#of 1)}-\nu  \right) + \frac{\nu}{2^L}\boldsymbol{1} \cdot \boldsymbol{1}^\dagger=D_1+ \frac{\nu}{2^L}\boldsymbol{1} \cdot \boldsymbol{1}^\dagger
\end{align}
where $\boldsymbol{1}$ is the vector with only 1s as entries. A trivial basis of this space is given by the $2^L$ vectors indexed by their ordered sequence of 0s and 1s, that we can note $e_{\boldsymbol{\sigma}}$, $\boldsymbol{\sigma}$ being a vector of size $L$ of 0s and 1s.\\
We search for eigenvalues of the $A_i$'s. Let us note $\mathcal{E}_k$ the subspace generated by vectors where $k$ nucleotides are $0$ (generated by the
$e_{\boldsymbol{\sigma}}$'s where $\sum_i \sigma_i=L-k$), and
$\mathcal{H}_k=\mathcal{E}_k \cap \lbrace v| v \cdot \boldsymbol{1}=0 \rbrace$.
We note that the dimension of $\mathcal{E}_k$ is $\binom{L}{k}$ while the one of $\mathcal{H}_k$ is $\binom{L}{k}-1$. Besides, all elements in $v$ in $\mathcal{H}_k$ are eigenvectors of both $A_0$ and $A_1$ of eigenvalues $k$ and $L-k$ respectively:
\begin{align}
     A_0 \cdot v &= D_0 \cdot v + \frac{\nu}{2^L}\boldsymbol{1} \cdot \boldsymbol{1}^\dagger \cdot v =k v \\
     A_1 \cdot v &= (L-k) v
\end{align}
Consequently, we found $2^L-(L+1)$ (independent) eigenvectors of $\exp\left(\tau A_1\right) \cdot \exp\left(\tau A_0\right)$ of same eigenvalue $e^{L\tau}$ as for any $k$ and $v \in \mathcal{H}_k$
\begin{align}
    \exp\left(\tau A_1\right) \cdot \exp\left(\tau A_0\right) \cdot v=e^{\tau (L-k)+\tau k} v=e^{L\tau }v \; .
\end{align}
Thus, $\Lambda_L$ is given by the maximum between $e^{L\tau }$ and the maximum eigenvalue on the projected space of dimension $L+1$ corresponding to the remaining eigenvectors to find. An orthonormal basis of this space is simply given by the $L+1$ vectors we will note $\lvert k \rangle$. Each $\lvert k \rangle$ corresponds to the sum of all $\binom{L}{k}$ basis vector in $\mathcal{E}_k$ (vectors with a 1 at the corresponding sequence with $k$ 0s, 0 for all other entries), normalised by $1/\sqrt{\binom{L}{k}}$, \\
\begin{align}
\lvert k \rangle=\frac{1}{\sqrt{\binom{L}{k}}}\sum_{\boldsymbol{\sigma}|\sum_i \sigma_i=L-k}e_{\boldsymbol{\sigma}} \; .
\end{align}
By projecting $A_0$ and $A_1$ in this basis, we get the projected $A_0$ and $A_1$ (we keep the same notation in the projected space):
\begin{align}
    A_0= \sum_{k=0}^{L} (k-\nu) \lvert k \rangle \langle k \lvert+\frac{\nu}{2^L}\sum_{k,\ell=0}^{L} \sqrt{\binom{L}{k}}\sqrt{\binom{L}{\ell}} \lvert k \rangle \langle \ell \lvert \\
    A_1= \sum_{k=0}^{L} (L-k-\nu) \lvert k \rangle \langle k \lvert+\frac{\nu}{2^L}\sum_{k,\ell=0}^{L} \sqrt{\binom{L}{k}}\sqrt{\binom{L}{\ell}} \lvert k \rangle \langle \ell \lvert
\end{align}
where we used that 
\begin{align}
    \boldsymbol{1} \cdot \boldsymbol{1}^\dagger  \lvert k \rangle = \sum_\ell \sqrt{\binom{L}{k}}\sqrt{\binom{L}{\ell}} \langle \ell \lvert
\end{align}
This first step gives us a much smaller matrix to diagonalize ($2^L$ versus $L+1$), and we can check our calculations numerically by computing the top eigenvalue for both matrices. \\
\begin{figure}[th!]
    \centering
    \includegraphics[width=0.8\columnwidth]{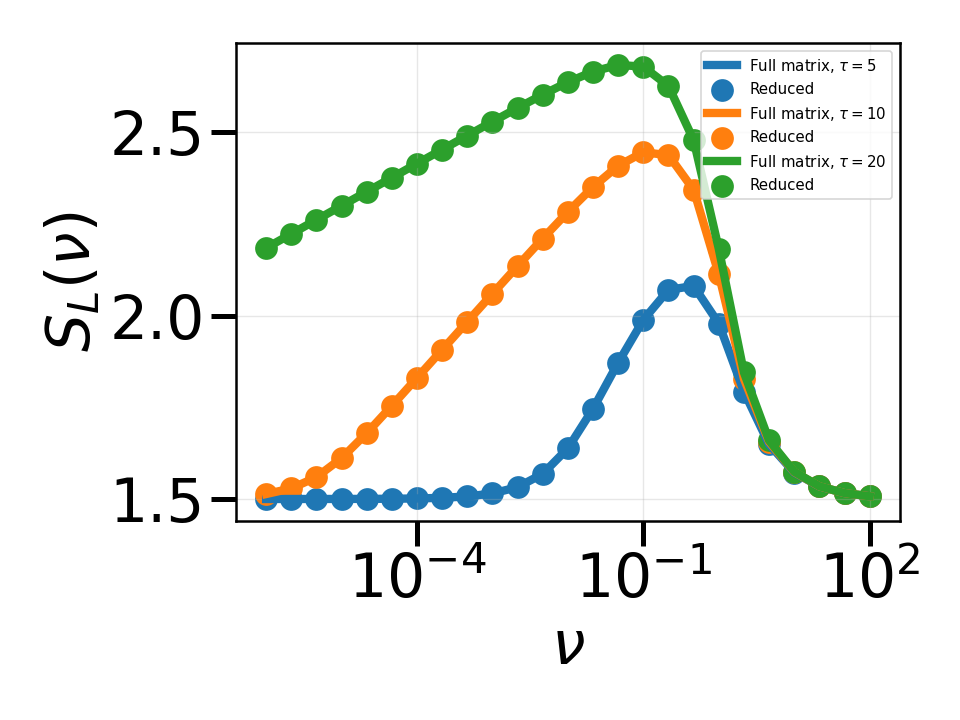}
    \caption{Comparison of the effective growth rate $S_L(\nu)$ obtained from the top eigenvalue of the $2^L$ matrix and the $L+1$ matrix for $L=3$.}
    \label{fig:Reduction}
\end{figure}
\newpage
\paragraph{Perturbation theory at small $\nu/2^L$}
Next, we would like to find an approximate expression of the eigenvalues and eigenvectors of both $A_0$ and $A_1$ in order to compute $\Lambda_L$.\\
For that, we use perturbation theory at first order in $\nu/2^L$, using that at $\nu=0$ both $A_0$ and $A_1$ are both diagonal in the basis of $\left( \lvert k \rangle \right)_{0\leq k \leq L}$, of non-degenerate eigenvalues. We note $\left( \lvert k^{(0)} \rangle \right)_{0\leq k \leq L}$ and $\left( \lambda_k^{(0)}  \right)_{0\leq k \leq L}$ the first order shifted eigenvectors and eigenvalues of $A_0$ (resp. $\left( \lvert k^{(1)} \rangle \right)_{0\leq k \leq L}$ and $\left(  \lambda_k^{(1)} \right)_{0\leq k \leq L}$ the first order shifted eigenvectors and eigenvalues of $A_1$). We get that:
\begin{align}
    \lvert k^{(0)} \rangle=\lvert k \rangle+\frac{\nu}{2^L}\sum_{\ell \neq k} \frac{\sqrt{\binom{L}{k}}\sqrt{\binom{L}{\ell}}}{k-\ell}\lvert \ell \rangle \\
     \lambda_k^{(0)} =k-\nu +\frac{\nu}{2^L}\binom{L}{k}
\end{align}
and 
\begin{align}
    \lvert k^{(1)} \rangle=\lvert k \rangle+\frac{\nu}{2^L}\sum_{\ell \neq k} \frac{\sqrt{\binom{L}{k}}\sqrt{\binom{L}{\ell}}}{\ell-k}\lvert \ell \rangle \\
     \lambda_k^{(1)} =L-k-\nu +\frac{\nu}{2^L}\binom{L}{k} \; .
\end{align}
Of note , the two basis of eigenvectors are two (distinct) orthonormal basis at first order in $\nu$. From this, we get the formal expression for $\exp\left(\tau A_1\right) \cdot \exp\left(\tau A_0\right)$:
\begin{align}
    \exp\left(\tau A_1\right) \cdot \exp\left(\tau A_0\right)=\sum_{k,\ell }e^{(\lambda_k^{(0)}+\lambda_\ell^{(1)})\tau}\lvert k^{(0)} \rangle \langle k^{(0)} \lvert \ell^{(1)}\rangle \langle \ell ^{(1)} \lvert
\end{align}
We can evaluate this expression on all the eigenvectors $\lvert k^{(0)} \rangle$, taking the maximal one as a proxy for $\Lambda_L$ (rigorously: a lower bound).
\begin{align}
   \langle k  ^{(0)} \lvert \exp\left(\tau A_1\right) \cdot \exp\left(\tau A_0\right) \lvert k  ^{(0)} \rangle &= e^{\lambda_k^{(0)}\tau} \sum_{\ell} e^{\lambda_\ell^{(1)}\tau} \langle k^{(0)} \lvert \ell^{(1)}\rangle^2  
\end{align}
The exponential terms are maximized for $k=L$ and $\ell=0$, which provides us with a (good) lower bound for $\Lambda_L$ (by neglecting all other exponential terms):
\begin{align}
    \Lambda_L \geq e^{\lambda_L^{(0)}\tau}  e^{\lambda_0^{(1)}\tau} \langle L^{(0)} \lvert 0^{(1)}\rangle^2=e^{\tau(2L -2\nu+2\nu/2^L) }\left(\frac{2\nu}{2^L L} \right)^2
\end{align}
Finally, we obtain the approximate lower-bound estimate for $S_L(\nu)$:
\begin{align}
    S_L(\nu) \geq \frac{\ln \max\left(e^{L\tau},e^{\tau(2L -2\nu+2\nu/2^L) }\left(\frac{2\nu}{2^L L} \right)^2\right) }{2\tau}=\max\left(\frac{L}{2}, L-\nu(1-2^{-L})+\frac{\ln \left(\frac{2\nu}{2^L L} \right)}{\tau} \right) \label{eq:Fitness_Q2}
\end{align}
We can find the optimal DGR switching rate $\nu^*$ by maximizing this expression over $\nu$ and get
\begin{align}
    \nu^*=\frac{1}{\tau(1-2^{-L})}
\end{align}
which confirms the optimality of taking $\nu\approx 1/\tau$ even in the DGR setting with length $L>1$.\\
We expect this lower bound to work better at:
\begin{itemize}
    \item $\nu \ll 1$: because it is perturbation theory in $\nu$. 
    \item $\frac{\nu}{2^L L}e^{\tau L} \geq 1$: the time for best sequence to appear and dominate the population. Otherwise, the exponentials are not dominant and do not dictate the term to keep in the sum. The interpretation is the following: the population currently selected needs some minimal time to grow and dominate by the end of the time window. Thus, the lower bound is poor at $\tau$, $\nu$ and $L$ too small.
\end{itemize}
\begin{figure}[th!]
    \centering
    \includegraphics[width=0.9\columnwidth]{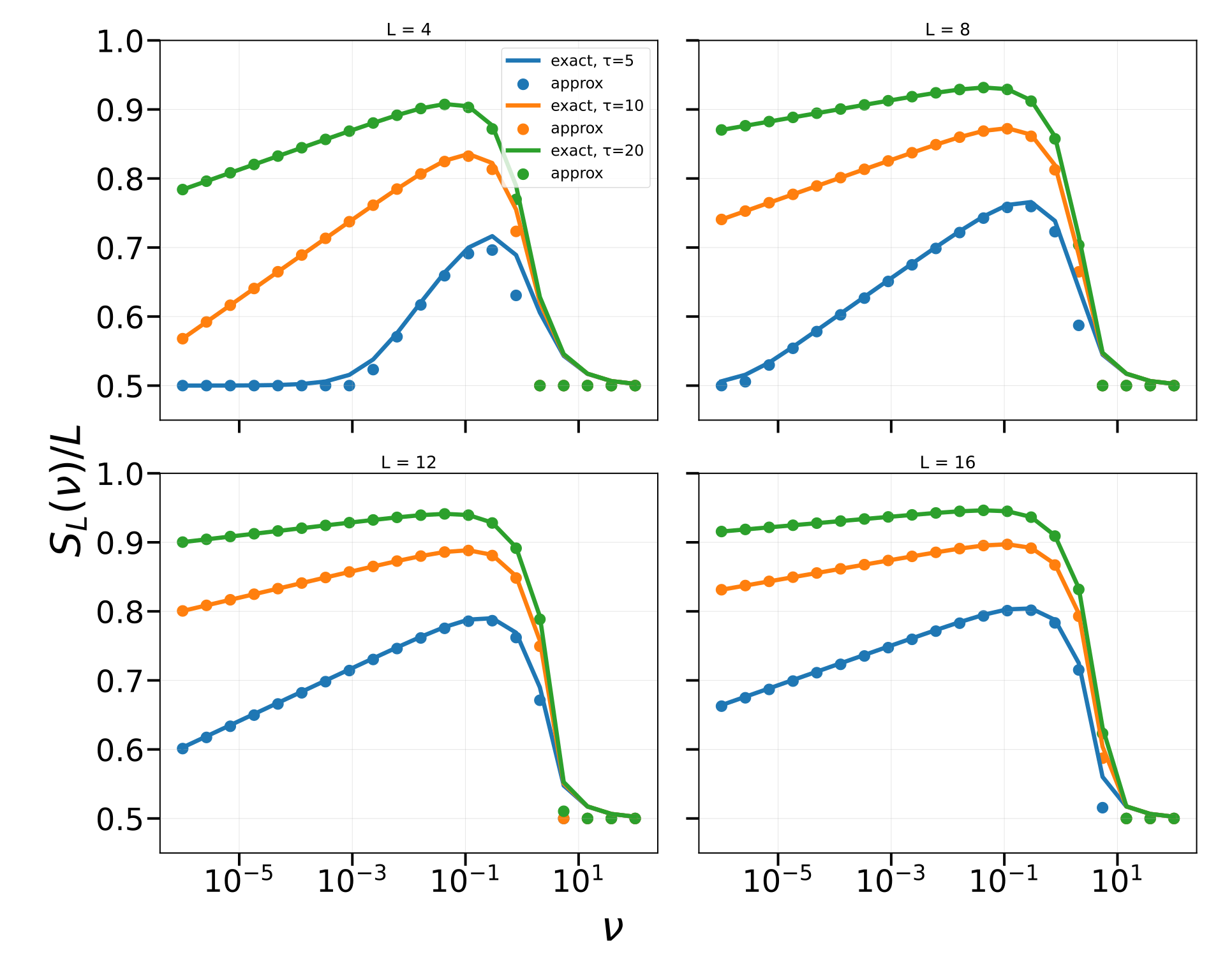}
    \caption{Lower bound of $S_L(\nu)$ versus its exact value by direct diagonalization of the $L+1$ matrix for $L=4$, $L=8$, $L=12$ and $L=16$.}
    \label{fig:Feff}
\end{figure}
\paragraph{Interpretation of the result}
At this point, we notice that the result Eq.~\eqref{eq:Fitness_Q2} can be derived in a simpler way. Let us consider an initial population $\boldsymbol{n}(0)$. During the first time interval $[0,\tau]$, the population with $L$ $0$s on their VR noted $n_{0\ldots0}$ grows exponentially at a rate $L-\nu+\nu/2^L$, and thus is the dominant population (the total population is approximatively given by $n_{0\ldots0}$). In particular, if we also consider the population $n_{1 \ldots 1}$ with $L$ $1$s in that time window, their dynamics obey
\begin{align}
     \dot{n}_{1\ldots1}(t) =\frac{\nu}{2^L} n_{0\ldots0}(t)
\end{align}
such that at time $\tau$,
\begin{align} \label{eq:CI_n1}
    n_{1\ldots1}(\tau)\approx \frac{ \nu } {L 2^L}n_{0\ldots 0}(\tau)
\end{align}
using that $n_{0\ldots 0}$ grows exponentially at rate $L-\nu+\nu/2^L\approx L$.\\
When the environment changes to select $1$s, we consider the population of $L$ $0$s in their {\bf VR} which switches to $L$ $1$s. Besides, we only consider this transition, neglecting all others (individuals which do other transitions are lost). The dynamics is then given by:
\begin{align}
    \dot{n}_{1\ldots 1}(t)=(L-\nu+\nu/2^L)n_{1\ldots 1}(t)+\frac{\nu}{2^L}n_{0 \ldots 0}(t) \; .
\end{align}
Because $n_{0 \ldots 0}(t)=n_{0 \ldots 0}(\tau)$ for $t>\tau$ (only $1$s are selected, $0$s have zero growth rate), we have that:
\begin{align}
    n_{1\ldots 1}(2 \tau)&=\frac{\nu}{2^L (L-\nu(1-1/2^L))}n_{0 \ldots 0}(\tau)\left(e^{(L-\nu+\nu/2^L)\tau}-1 \right)+n_{1\ldots 1}(\tau)e^{(L-\nu+\nu/2^L)\tau} \\
    &\approx e^{(L-\nu+\nu/2^L)\tau}\left( \frac{\nu}{2^L L} n_{0 \ldots 0}(\tau)+ n_{1 \ldots 1}(\tau)\right) \; \\
    & \approx e^{(L-\nu+\nu/2^L)\tau}\frac{2\nu}{2^L L} n_{0 \ldots 0}(\tau) \; .
\end{align}
To summarize, during each environment window, the total population will grow by a factor $\frac{2\nu}{2^L L}e^{(L-\nu+\nu/2^L)\tau}$ at least thanks to the transition $0 \ldots 0$ to $1 \ldots 1$. Thus, the effective growth rate is at least 
\begin{align}
    S_L(\nu)\geq \frac{\ln \left( \frac{2\nu}{2^L L}e^{(L-\nu+\nu/2^L)\tau} \right)}{\tau}
\end{align}
which gives exactly Eq.~\eqref{eq:Fitness_Q2} obtained via perturbation theory (the other lower bound $L/2$ coming simply from a population without any DGR switching which grows of a factor $L\tau$ during one bout and stays fixed during the next one). Thus, it means that, to first order in $\nu$, the main contribution of the DGR mechanism to the fitness is given by this transition of the fittest sequence of the previous bout to the next fittest sequence. We will use that directly when considering the case $Q>2$. 

\subsubsection{General $Q$ lower bound}
\begin{figure}[th!]
    \centering
    \includegraphics[width=0.6\columnwidth]{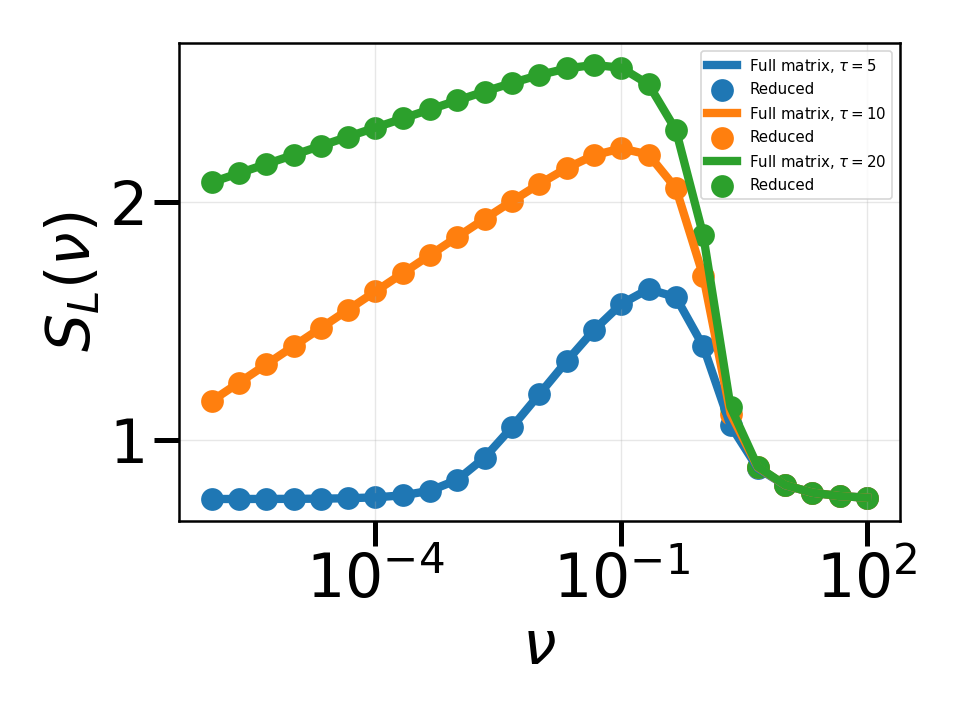}
    \caption{Comparison of the top eigenvalue of the $Q^L$ matrix and the $O(L^3)$ matrix for $L=3$.}
    \label{fig:Reduction_Q4}
\end{figure}

\paragraph{Reduction of the matrix size.}
We note $A_\kappa$ the rate matrix for state $\kappa$,
\begin{align}
    A_\kappa=\text{diag}\left(s \text{(\# of }\kappa)-\nu  \right) + \frac{\nu}{2^L}\boldsymbol{1} \cdot \boldsymbol{1}^\dagger \; .
\end{align}
We can apply the same projection procedure to the $A_\kappa$ matrices to find the highest eigenvalue in the smaller vector space. For any $\kappa$ we have the $A_\kappa$ matrix (similarly to the previous part) for $\kappa=0, \ldots, Q-1$:
\begin{align}
    A_{\kappa}&= \sum_{c_0,\ldots,c_{Q-1}|\sum_k c_k=L} (c_{\kappa}-\nu) \lvert c_0,\ldots,c_{Q-1} \rangle \langle c_0,\ldots,c_{Q-1} \lvert \nonumber
    \\&+\frac{\nu}{Q^L}\sum_{c_0,\ldots,c_{Q-1},c'_0,\ldots,c'_{Q-1}|\sum_k c_k=L, \sum_k c'_k=L}^{L} \sqrt{\binom{L}{c_0,\ldots,c_{Q-1}}}\sqrt{\binom{L}{c_0,\ldots,c_{Q-1}}} \lvert c_0,\ldots,c_{Q-1} \rangle \langle c'_0,\ldots,c'_{Q-1} \lvert
\end{align}
where $c_k$ is the number of symbol $k$ in the sequence. We see indeed that this is exact in \ref{fig:Reduction_Q4}, meaning that we can use this smaller matrix to compare our approximated results that we will obtain in the following section.\\

\begin{figure}[th!]
    \centering
    \includegraphics[width=0.7\columnwidth]{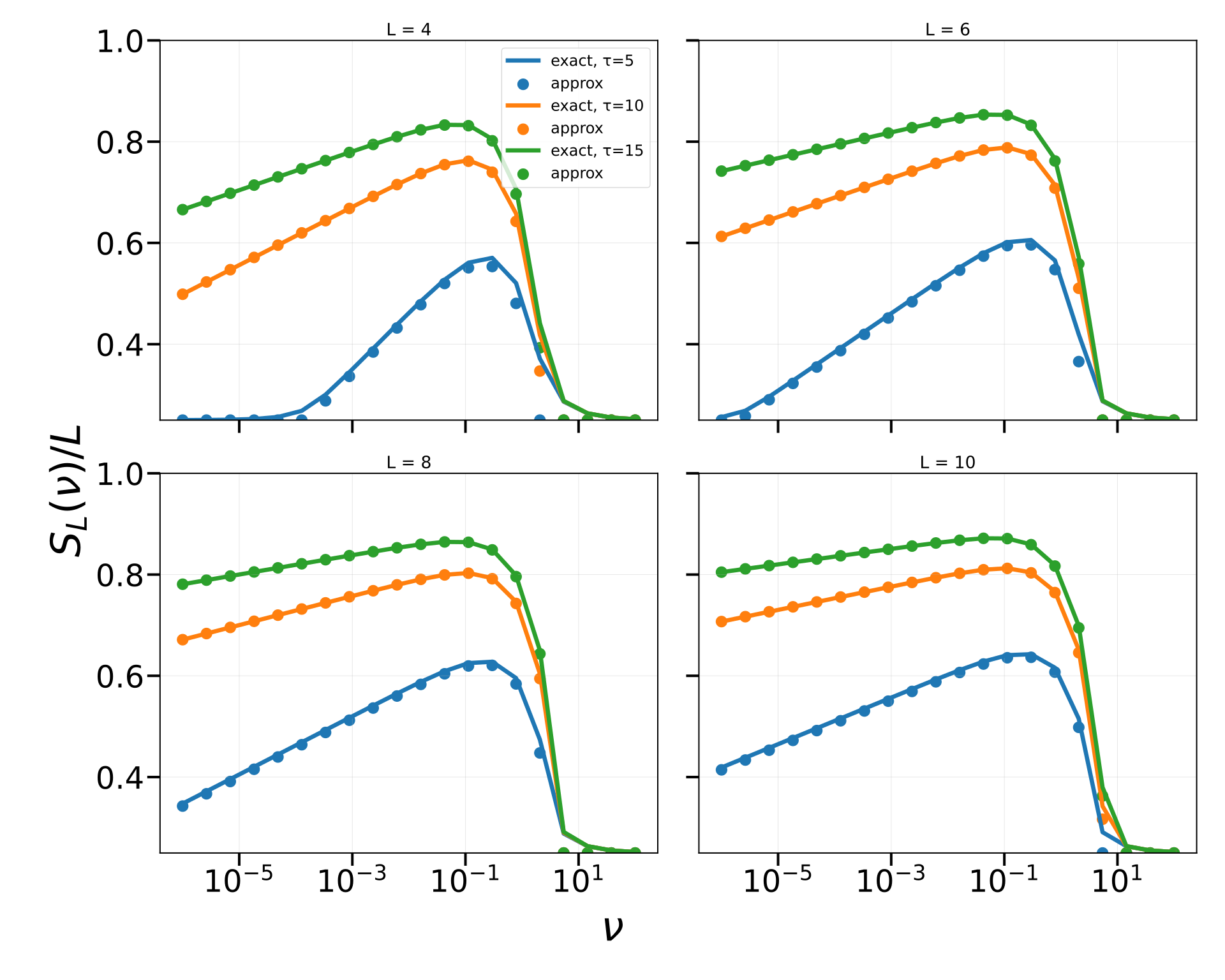}
    \caption{Lower bound of $S_L(\nu)/L$ versus its exact value by direct diagonalization of the $O(L^3)$ square matrix for $L=4$, $L=6$, $L=8$ and $L=10$.}
    \label{fig:Feff_Q4}
\end{figure}

\paragraph{Effective fitness expression.}
A lower bound is given by
\begin{align}
    S_L(\nu)\geq \frac{\ln \max\left(e^{L\tau},e^{Q\tau(L -\nu+\nu/Q^L) }\left(\frac{2\nu}{Q^L L} \right)^Q\right) }{Q\tau}=\max\left(\frac{L}{Q}, L-\nu(1-Q^{-L})+\frac{\ln \left(\frac{2\nu}{Q^L L} \right)}{\tau} \right) \; .
\end{align}
It corresponds to the path of states $0\ldots 0$ to $1\ldots 1$ to $2\ldots2$, so on and so forth where transition from one state to the other happens at rate $\nu/Q^L$. Indeed, because the growth rate is $L$ in the new state, we get that the population $n_1$ in the newly selected state at time $t$ knowing the population in the previously selected state $n_0$  (which will not grow anymore as it is not selected anymore) obeys the equation:
\begin{align}
    \frac{{\rm d}n_1}{{\rm d}t}=(L-\nu(1-1/Q^L))n+\frac{\nu}{Q^L}n_0
\end{align}
of solution 
\begin{align}
n_1(\tau)\approx e^{(L-\nu(1-1/Q^L))\tau}\left( \frac{\nu}{L Q^L } n_0(0)+n_1(0) \right) 
\end{align}
at large $\tau$. Using that during the previous environment the population $n_0$ grew exponentially at rate $L$ and mutated at rate $\nu/Q^L$ into the population $n_1$, we get as in \eqref{eq:CI_n1} that
\begin{align}
    n_1(0)=\frac{\nu}{LQ^L}n_0(0) \; .
\end{align}
which leads to 
\begin{align}
n_1(\tau)\approx e^{(L-\nu(1-1/Q^L))\tau}\frac{2\nu}{Q^L L} n_0(0) \; .
\end{align}
Doing the operation $Q$ times gives us the lower bound above, that we check in \ref{fig:Feff_Q4}. This gives us back $\nu^*=\frac{1}{\tau(1-Q^{-L})}$.

\begin{figure}[th!]
    \centering
    \includegraphics[width=0.8\columnwidth]{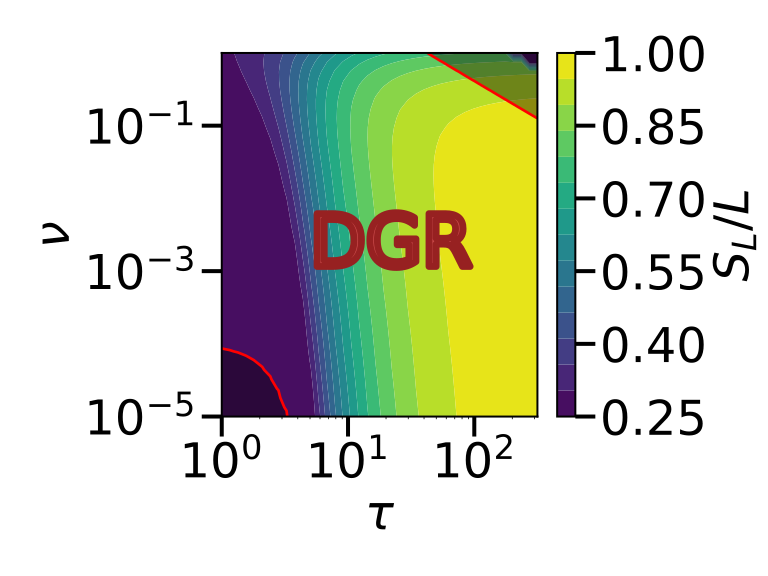}
    \caption{Phase diagram of the DGR vs non-DGR system as a function of $\nu$ and $\tau$, for $\mu=10^{-4}$ and $L=5$. The region where  $S_L(\nu)>LS_1(\mu)$ is delimited by the red contour. The region $S_L(\nu)<LS_1(\mu)$ is darkened.}
    \label{fig:PD_tau_L}
\end{figure}

We also show in \ref{fig:PD_tau_L} the phase diagram in $(\tau,\nu)$ which delimits the parameter space region where the DGR system dominates the non DGR one {\it i.e.} $S_L(\nu)>LS_1(\mu)$.

\clearpage

\section{Slow dynamics of the TR}
\subsection{$L=1$}
To describe the {\bf TR} dynamics, we will separate the discussion by separating over the diverse type of {\bf (TR,VR)} genotypes. We note $A$ the nucleotide which corresponds to the diversification (when present in the {\bf TR}), $T$ corresponding to the currently selected nucleotide (when present in the {\bf VR}), and $N$ for all the rest (we note $M$ when we include $A$ in this population). We represent the graph of population fluxes between the different {\bf (TR, VR) } populations.  \\
\begin{figure}[th!]
    \centering
    \includegraphics[width=0.5\columnwidth]{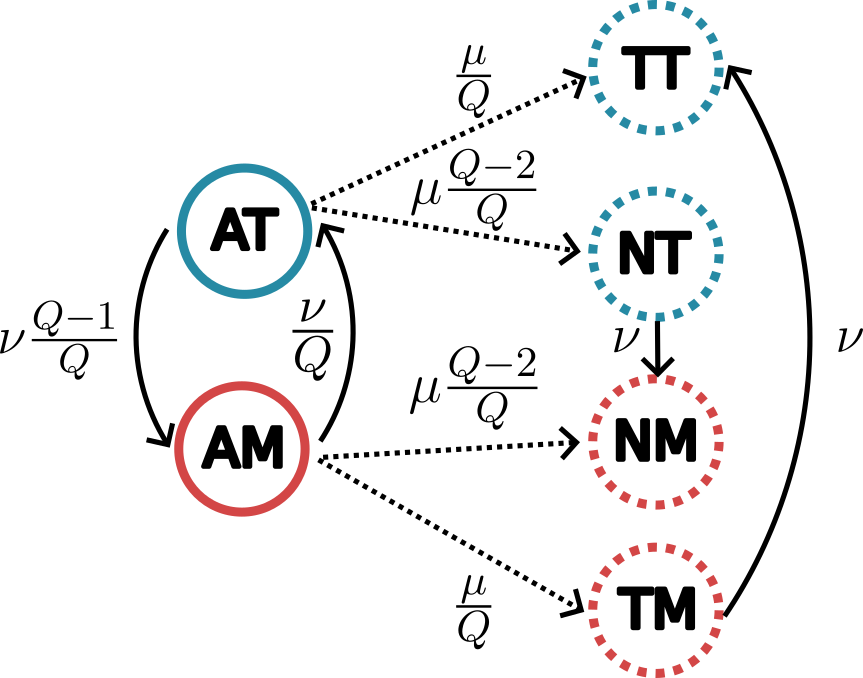}
    \caption{{\bf The graph of population transitions.} The graph of effective transitions between the different {\bf (TR,VR)} population (first letter is the {\bf TR}, second the {\bf VR}). $A$ corresponds to the diversifying nucleotide, $T$ to the currently selected nucleotide, $M$ to any nucleotide except form $T$, $N$ to any nucleotide except from $A$ and $T$. In blue are the population with growth rate $1$, in red those with growth rate $0$. Hashed circles correspond to population of size of order $\mu$, while plain lines are of order $1$. Dotted arrows represent transitions of order $\mu$, plain line arrows those of order $\nu$. We do not represent transitions which would lead to $\mu^2$ corrections to the population sizes.}
    \label{fig:PopArrows}
\end{figure}
Of note, we make the assumption that $T\ne A$, hypothesis we discuss in the end of the section.
\subsubsection{The {\bf TR}=A population: first order}
\begin{figure}[th!]
    \centering
    \includegraphics[width=0.8\columnwidth]{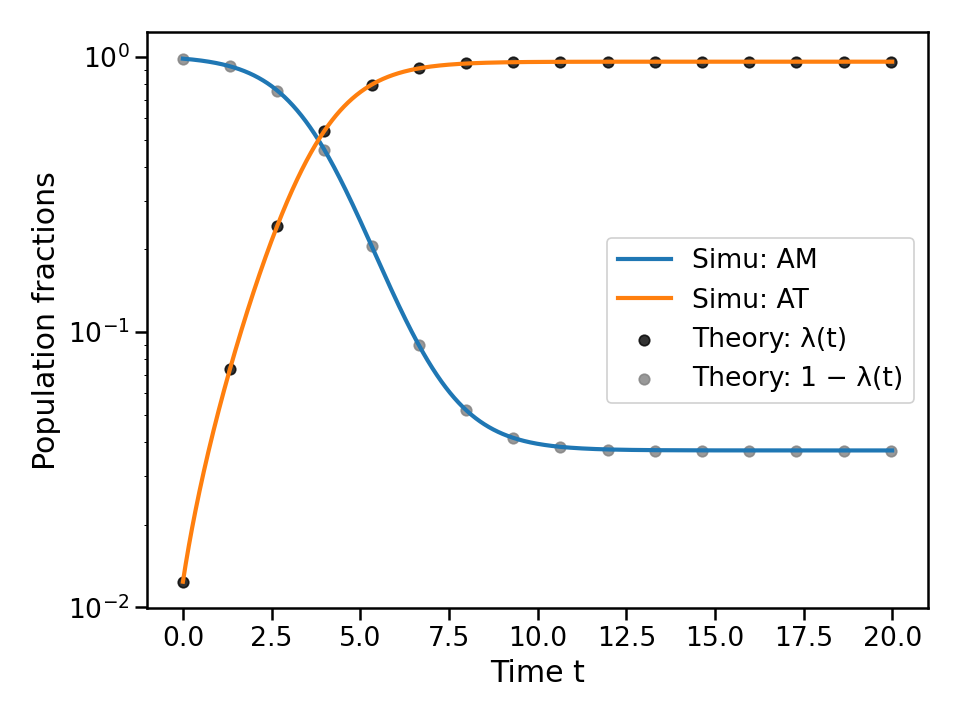}
    \caption{{\bf A population fraction.} Comparison between numerical simulations (lines) and theoretical predictions (dots) of the population of {\bf TR} having an adenine. The population size is $10^7$, and the parameters are $\tau=20$, $\nu=5.10^{-2}$, $\mu=10^{-5}$, $Q=4$.}
    \label{fig:A_pop}
\end{figure}

As seen in the first part, by taking $\nu\approx 1/\tau$ we are going to select for {\bf TR} with the Adenine. Thus, as a first step, we describe this population, which should dominate the whole population. To do so, we split this population in two: either they have the right {\bf VR}=T (population of size $n_{AT}(t)$) or they don't (population of size $n_{AM}(t)$). The equations describing the joint dynamics of these two populations are given by: 
\begin{align}
        \dot{n}_{AT}&=\left(1-\nu\frac{Q-1}{Q}\right)n_{AT}+\frac{\nu}{Q}n_{AM}\\
    \dot{n}_{AM}&=-\frac{\nu}{Q} n_{AM}+\nu\frac{Q-1}{Q}n_{AT}
\end{align}
of solution given by:
\begin{align}
    n_{AT}(t)&=C_+ e^{\lambda_+ t} +C_- e^{\lambda_- t}\\
    n_{AM}(t)&=C_+ x_+ e^{\lambda_+ t} +C_- x_- e^{\lambda_- t}
\end{align}
where
\begin{align}
\Delta = \sqrt{(1-\nu)^{2} + \frac{4\nu}{Q}}, \qquad
\lambda_{\pm} = \frac{1-\nu \pm \Delta}{2}.
\end{align}
and 
\begin{align}
    x_\pm=\frac{Q\lambda_\pm-Q+ \nu(Q-1)}{\nu}\\
    C_+=\frac{n_{AM}(0)-x_+ n_{AT}(0)}{x_+-x_-}\\
    C_-=\frac{x_+ n_{AT}(0)-n_{AM}(0)}{x_+-x_-}
\end{align}
For the initial conditions, we use that at times $t\gg 1/(\lambda_+-\lambda_-)=1/\Delta$,
\begin{align}
    \frac{n_{AM}(t)}{n_{AT}(t)+n_{AM}(t)}\sim \frac{x_+}{1+x_+} \, .
\end{align}
However, we know that the initial fraction of the population in $(A,T)$ and $(A,M)$ are given by the respective fractions at the previous time window, where the $(A,T)$ population corresponds to a fraction $1/(Q-1)$ of the previous $(A,M)$ population ($T$ not being the selected nucleotide in the previous time window and supposing that the behaviour of all non positively selected nucleotide is the same).
Thus, the initial fraction of sites starting in state {\bf (TR,VR)}$=$(A,T) can be approximated by
\begin{align}
    n_{AT}(0)=\frac{x_+}{(Q-1)(1-x_+)} \, , \, \, n_{AM}(0)=1-n_{AT}(0)
\end{align}
From this, we get the fraction of the population in each state,
\begin{align}
    f_{AT}(t)=\frac{n_{AT}(t)}{n_{AM}(t)+n_{AT}(t)} \, , \, \, f_{AM}(t)=\frac{n_{AM}(t)}{n_{AM}(t)+n_{AT}(t)} \; .
\end{align}
Using this, we get the full description of the {\bf TR}$=$A population as well as the average fitness $\lambda(t)$ which is given by the fraction of the population in the $AT$ state: 
\begin{align}
\lambda (t)= 1 \times f_{AT}(t) +0\times f_{AM}(t)=f_{AT}(t)
\end{align}
We check our analytical results against  numerical simulations in \ref{fig:A_pop}.

\newpage 

\subsubsection{The {\bf TR}$\neq$A population: the order $\mu$ correction}
To describe this population, we directly compute the fraction dynamics using Wright-Fisher equations \cite{Neher:2011} as we computed the average fitness dynamics $\lambda(t)$ (to first order, without the $\mu$ corrections). In the following, we can replace when necessary $f_{AT}(t)=\lambda(t)$ and $f_{AM}(t)=1-\lambda(t)$. We will note $\Lambda(t)\equiv\int_0^t \lambda(s) {\rm d}s$.

\paragraph{{\bf TR}$=$T,  {\bf VR}$=$ M}
\begin{align}
   \dot{f}_{TM}(t)=\left(-\nu-\lambda (t)\right)f_{TM}(t) +\frac{\mu}{Q}f_{AM}(t)
\end{align}
such that
\begin{align}
    f_{TM}(t)&=f_{TM}(0)e^{-\nu t -\Lambda(t)}+\frac{\mu}{Q}e^{-\nu t -\Lambda(t)}\int_0^t e^{\nu s +\Lambda(s)}f_{AM}(s){\rm d}s \\
    &=f_{TM}(0)e^{-\nu t -\Lambda(t)}+\frac{\mu}{Q}e^{-\nu t -\Lambda(t)}\int_0^t e^{\nu s +\Lambda(s)}(1-\lambda(s)){\rm d}s
\end{align}

\paragraph{{\bf TR}$=$T,  {\bf VR}$=$T}
\begin{align}
   \dot{f}_{TT}(t)=(1-\lambda (t))f_{TT}(t) + \frac{\mu}{Q}f_{AT}(t)+\nu f_{TM}(t)
\end{align}
such that
\begin{align}
    f_{TT}(t)&=f_{TT}(0)e^{t-\Lambda(t)}+e^{t-\Lambda(t)}\int_0^t \left( \frac{\mu}{Q}f_{AT}(s)+\nu f_{TM}(s) \right)e^{-s+\Lambda(s)}{\rm d}s\\
    &=f_{TT}(0)e^{t-\Lambda(t)}+e^{t-\Lambda(t)}\int_0^t \left( \frac{\mu}{Q}\lambda(s)+\nu f_{TM}(s) \right)e^{-s+\Lambda(s)}{\rm d}s
\end{align}
\paragraph{{\bf TR}$=$N, {\bf VR}$=$T}
\begin{align}
   \dot{f}_{NT}(t)=(1-\lambda (t)-\nu)f_{NT}(t) + \frac{\mu(Q-2)}{Q}\left(   f_{AT}(t)+f_{TT}(t)\right)
\end{align}
We neglect $f_{TT}(t) \ll f_{AT}(t)$:
\begin{align}
    f_{NT}(t)&=f_{NT}(0)e^{t-\Lambda(t)-\nu t}+\frac{\mu(Q-2)}{Q}e^{t-\Lambda(t)-\nu t} \int_{0}^t e^{-s+\Lambda(s)+\nu s}f_{AT}(s) {\rm d}s \\
    &=f_{NT}(0)e^{t-\Lambda(t)-\nu t}+\frac{\mu(Q-2)}{Q}e^{t-\Lambda(t)-\nu t} \int_{0}^t e^{-s+\Lambda(s)+\nu s}\lambda(s) {\rm d}s
\end{align}
\paragraph{{\bf TR}$=N$, {\bf VR}$=$M}
\begin{align}
   \dot{f}_{NM}(t)=-\lambda (t)f_{NM}(t) + \frac{\mu(Q-2)}{Q}f_{AM}(t)+\nu f_{N T}(t)
\end{align}
of solution given by
\begin{align}
    f_{NM}(t)&=f_{NM}(0)e^{-\Lambda(t)}+e^{-\Lambda(t)} \int_{0}^t e^{\Lambda(s)}\left(\frac{\mu(Q-2)}{Q}f_{AM}(s)+\nu f_{NT}(s) \right){\rm d}s \\
    &=f_{NM}(0)e^{-\Lambda(t)}+e^{-\Lambda(t)} \int_{0}^t e^{\Lambda(s)}\left(\frac{\mu(Q-2)}{Q}(1-\lambda(s))+\nu f_{NT}(s) \right){\rm d}s
\end{align}

We then have to solve the self consistent equation which connects the different fractions together:
\begin{align} \label{eq:CI_nonA_1}
    f_{TM}(0)&=\frac{f_{NT}(\tau)}{Q-2} \\
    f_{TT}(0)&=\frac{f_{NM}(\tau)}{Q-2} \\
    f_{NT}(0)&=f_{TM}(\tau)\frac{Q-2}{Q-1}\\
    f_{NM}(0)&=f_{TT}(\tau)+f_ {TM}(\tau)\frac{Q-2}{Q-1}+f_{NM}(\tau)\frac{1}{Q-2}+f_{NT}(\tau)\frac{1}{Q-2} \; . \label{eq:CI_nonA_2}
\end{align}
The idea behind the self consistent equations is to consider that, at the next time epoch, it will be a different nucleotide (let's say G) that will be selected. We have to consider, when the selected nucleotide changes, where are the different populations going to be splitted. 
\begin{itemize}
    \item TT: going to NM population (not the best nucleotide in new time window).
    \item NT: a fraction $1/(Q-2)$ of the population is GT (N can be any of the $Q-2$ nucleotides) and thus goes to the GM population. The rest is going to the NM population (neither the selected nucleotide at {\bf TR} or at {\bf VR}).
    \item TM: a fraction $1/(Q-1)$ of the population is TG and goes to the NG population. The rest goes to the NM population. 
    \item NM: Because of the rapid replacement by the DGR activity, this population is mostly constituted in equal fractions of $({\bf TR}, {\bf VR})$ population where the {\bf TR} matches the {\bf VR}. In that case, there are effectively $Q-2$ subpopulations, with one corresponding to the GG population (thus a fraction $1/(Q-2)$). The rest goes to the $NM$ population. The other part of the population where the {\bf TR} does not match the {\bf VR} behaves similarly as the $TM$ population, in particular all the populations where the {\bf VR} is a G and the {\bf TR} is not an $A$, a $T$ or a $G$ (there are $Q-3$ of those): these will be added to the NG population. 
\end{itemize}

At this point, we observe as in the first part of the main text that the {\bf TR}$\neq$A population dynamics can be expressed as a linear problem: indeed, the time evolution between $[0,\tau[$ can be expressed as a linear transformation of the frequencies vector $\vec{f} \to M\vec{f}+\vec{b}$ (the expressions at time $\tau$ being affine in those at time $0$), while the final transformation of the subpopulations when the optimum sequence changes is of the form $\vec{f}\to C \vec{f}$. Thus, to find the initial conditions self-consistently, we have to find the solution $\vec{f}^*$ of the matrix problem $\vec{f}=C\cdot M \vec{f}+C\vec{b}$ .\\

\begin{figure}
    \centering
    \includegraphics[width=\columnwidth]{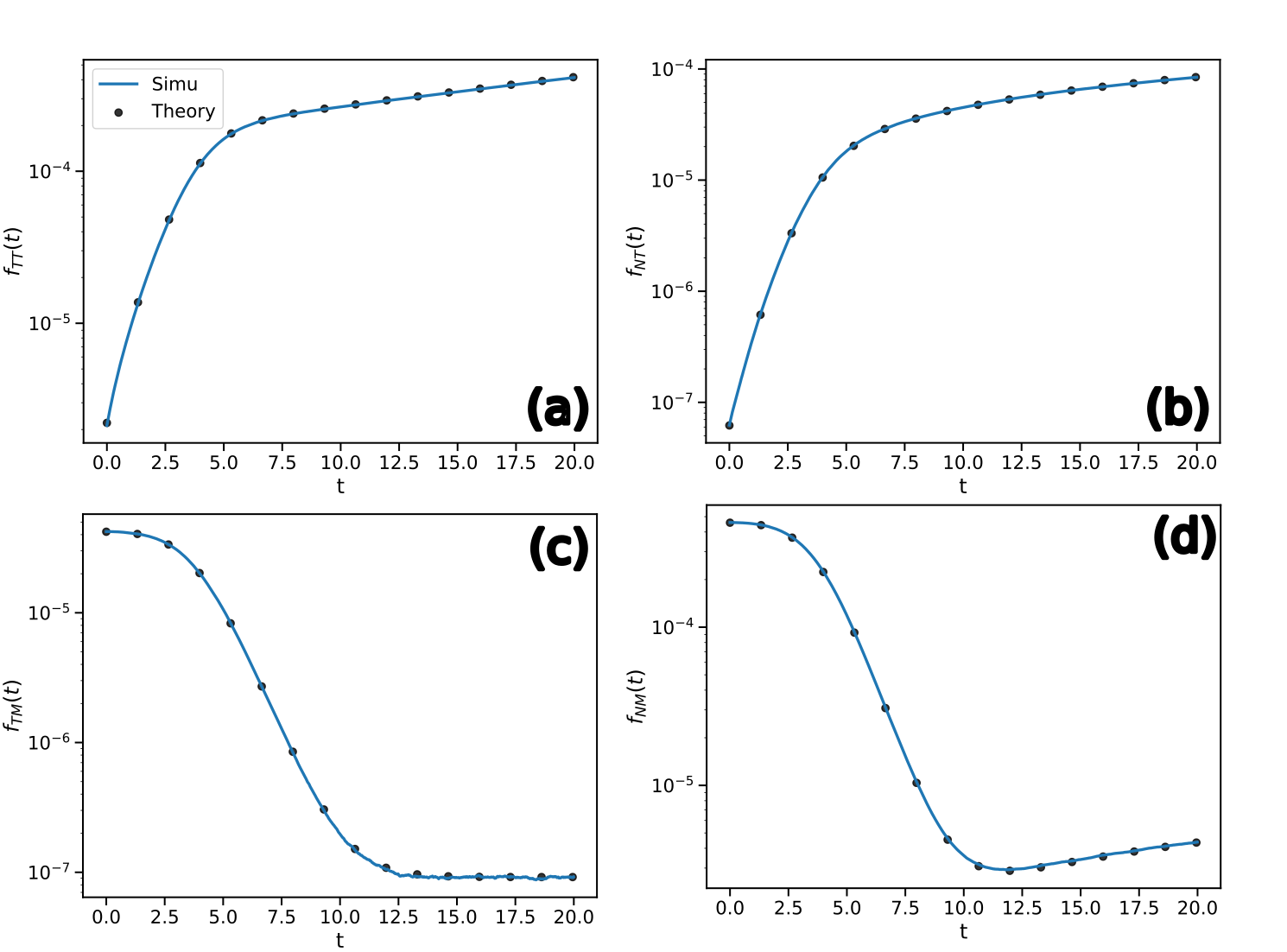}
    \caption{{\bf Non A populations fraction.} Comparison between numerical simulations (lines) and theoretical predictions (dots) of the fraction of the population of {\bf TR} not having an adenine. \textbf{(a)} The {\bf TR}$=$T, {\bf VR}$=$T population, \textbf{(b)} The {\bf TR}$=$T, {\bf VR}$=$M population,\textbf{(c)} The {\bf TR}$=$N, {\bf VR}$=$T population,\textbf{(d)} The {\bf TR}$=$N, {\bf VR}$=$M population. The population size is $10^{12}$, and the parameters are $\tau=20$, $\nu=5.10^{-2}$, $\mu=10^{-5}$, $Q=4$.}
    \label{fig:predict_subpop}
\end{figure}

We check that Eqs.~\eqref{eq:CI_nonA_1} to \eqref{eq:CI_nonA_2} allow to  correctly predict the dynamics of each population. In \ref{fig:predict_subpop}, we use the solution $\vec{f}^*=(I-C\cdot M)^{-1}\cdot C \vec{b}$ at time $\tau$ as initial condition for the results of subsections 1 to 4. We observe a perfect match between our simulations during the time window $[0,\tau]$ and the analytical results. Besides, we confirm that the $TT$ population dominates the population (of {\bf TR}$\neq$A) early in the dynamics, as supposed in the next section text.  

\begin{figure}
    \centering
    \includegraphics[width=0.8\columnwidth]{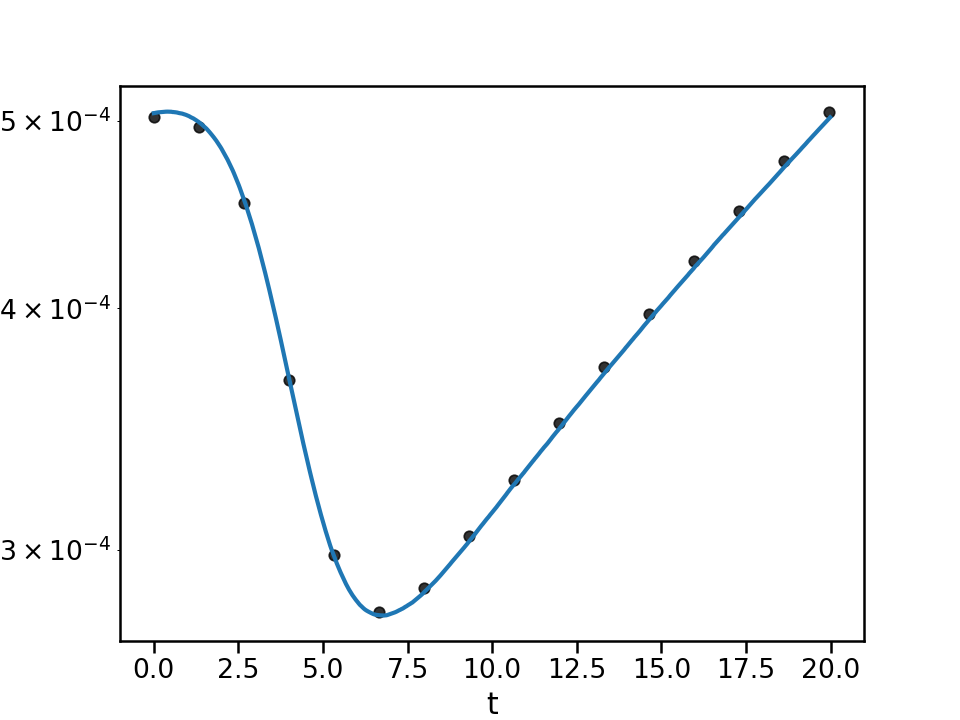}
    \caption{{\bf Non A population fraction.} Comparison between numerical simulations (lines) and theoretical predictions (dots) of the fraction of the population of {\bf TR} not having an adenine. The population size is $10^{12}$, and the parameters are $\tau=20$, $\nu=5.10^{-2}$, $\mu=10^{-5}$, $Q=4$.}
    \label{fig:not_A_pop}
\end{figure}

From these, we can evaluate the fraction of the population which has for {\bf TR} a nucleotide which is not $A$ as a function of time, displayed in \ref{fig:not_A_pop}.
Finally, we can evaluate the average fraction $f_\text{not A}$ sequences as a function of the parameters of the model and compare it to the results of numerical simulations. Once again, we obtain a perfect match with our analytical result.
\begin{figure}
    \centering
    \includegraphics[width=0.8\columnwidth]{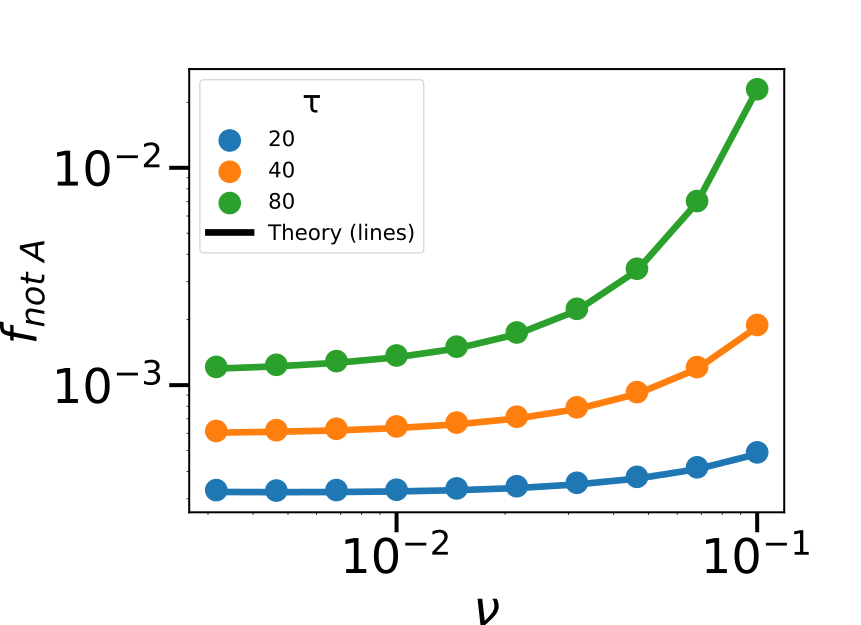}
    \caption{{\bf Non A population fraction.} Comparison between numerical simulations (lines) and theoretical predictions (dots) of the fraction of the population of {\bf TR} not having an adenine. The population size is $10^{12}$, and the parameters are $\mu=10^{-5}$, $Q=4$.}
    \label{fig:not_A_pop_exact}
\end{figure}
\clearpage
\subsubsection{Simplified dynamics}
We offer to do a simple approximation here to better understand the dynamics of the {\bf TR} population. The idea is that we consider the population with an Adenine in the {\bf TR} as the dominant one. As in the previous sections, we note $f_{AT}(t)$ (resp. $f_{AM}(t)$) the fraction of the population with an $A$ at the {\bf TR} and the currently selected nucleotide (resp. any non-selected nucleotide) noted $T$ (resp. $M$) in the {\bf VR}, with $T\neq A$ otherwise $A$ is surely selected in both the {\bf TR} and the {\bf VR}. The fitness advantage of having $T$ at the {\bf VR} is $s$. In the following, we fix time scales by taking $s=1$ (which is equivalent as taking the inverse growth rate as the unit of time). We use Wright-Fisher equations \cite{Neher:2011} to describe the population dynamics and note $\lambda(t)=1\times f_{AT}(t)+0\times f_{AM}(t)$ the average population fitness. At first, because $\nu \gg \mu$, we neglect the {\bf TR} mutation rate $\mu$ for the determination of the fraction of the population in each state. We get the following coupled equations: 
\begin{align} \label{eq:pop_A_dyn}
\begin{cases}
    \dot{f}_{AT}(t)&=(1-\lambda(t)-\nu \frac{Q-1}{Q})f_ {AT}(t)+\frac{\nu}{Q}f_ {AM}(t) \\
    \dot{f}_{AM}(t)&=(-\lambda(t)- \frac{\nu}{Q})f_ {AM}(t)+\nu\frac{Q-1}{Q}f_ {AT}(t)
\end{cases}
\end{align}
where $f_{AT}(t)+f_{AM}(t)=1$. We obtain the asymptotic fitness $\lambda(t) \to \lambda$ by solving Eq.~\eqref{eq:pop_A_dyn} when time derivatives are $0$ (and the condition that frequencies sum to 1). We obtain an asymptotic fitness (that we can approximate in the limit $\nu\ll 1$):
\begin{align}
    \lambda=\frac{1-\nu + \sqrt{(1-\nu)^{2} + \frac{4\nu}{Q}}}{2}\sim 1-\frac{Q-1}{Q}\nu
    \label{eq:lambda}
\end{align}
Then, to estimate the fraction of sequences which are not $A$ at the {\bf TR}, we consider that the main contributor corresponds to sequences from the $AT$ and $AM$ populations which mutated to the {\bf TR}=T population (which happens at rate $\mu/Q$). This hypothesis is well justified from the exact derivation and simulations of \ref{fig:predict_subpop}. Because the rate $\nu$ is much larger than $\mu$, we make the approximation that the {\bf VR} is immediatly replaced by a $T$ as well. This results in the following dynamics for the {\bf TR}$=$T, {\bf VR}$=$T population fraction, the population which lost its A in its {\bf TR} (in this approximation):
\begin{align}
    \dot{f}_{TT}(t)=(1-\lambda)f_{TT}(t)+\frac{\mu}{Q}
\end{align}
of solution, using Eq.~(\ref{eq:lambda}), given by:
\begin{align}
    f_{TT}(t)&=\frac{\mu}{Q(1-\lambda)}\left(e^{(1-\lambda)t}-1 \right) \nonumber \\
    &\sim \frac{\mu}{(Q-1)\nu}\left(e^{\frac{Q-1}{Q}\nu t }-1\right)
\end{align}
such that the average fraction of population without an A in their {\bf TR} is approximately given by:
\begin{align}
    f_\text{not A}&=\frac{1}{\tau}\int_0^\tau f_{TT}(t)  {\rm d}t \nonumber \\
    &=\frac{\mu}{\nu (Q-1)} \left[ \frac{Q}{(Q-1)\nu \tau}(e^{\frac{Q-1}{Q}\nu \tau}-1)-1\right] \; ,
\end{align}
as given in Eq.~(5) of the main text.

\subsubsection{Dicussion about $\bf VR$=A selection by the environment}
We discuss the case when the environment selects for Adenines in the {\bf VR}. \\
\begin{figure}[th!]
    \centering
    \includegraphics[width=0.5\columnwidth]{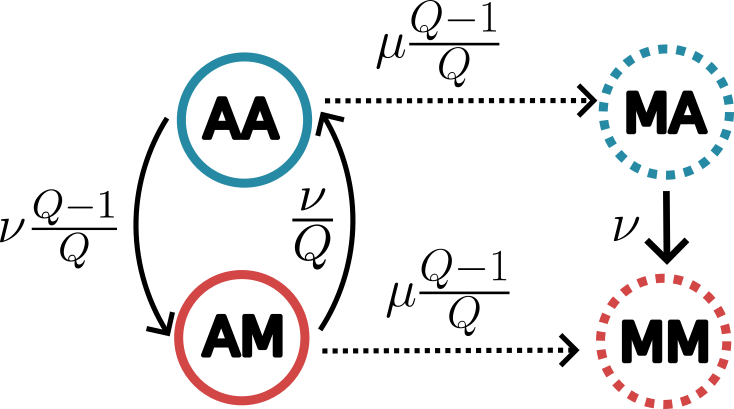}
    \caption{{\bf The graph of population transitions with adenine selection in the VR.} The graph of effective transitions between the different {\bf (TR,VR)} population (first letter is the {\bf TR}, second the {\bf VR}). $A$ corresponds to the diversifying nucleotide, $M$ to any nucleotide except form $A$. In blue are the population with growth rate $1$, in red those with growth rate $0$. Hashed circles correspond to population of size of order $\mu$, while plain lines are of order $1$. Dotted arrows represent transitions of order $\mu$, plain line arrows those of order $\nu$. We do not represent transitions which would lead to $\mu^2$ corrections to the population sizes.}
    \label{fig:PopArrows_A}
\end{figure}
As seen in the graph of \ref{fig:PopArrows_A} for transitions between populations, the dynamics of the population where {\bf TR}=A remain unchanged when the environment selects for an Adenine. Consequently, the asymptotic fitness remains
\begin{align}
\lambda \sim 1-\frac{Q-1}{Q}\nu \; .
\end{align}
When considering the population where {\bf TR} $\neq A$, the dynamics of the frequencies are governed by
\begin{align}
\dot{f}_{MA}(t) &= \left(1-\lambda -\nu\right)f_{MA}(t) + \mu \frac{Q-1}{Q}f_{AA}(t) \nonumber \\
\dot{f}_{MM}(t) &= -\lambda f_{MM}(t) + \mu \frac{Q-1}{Q}f_{AM}(t) \; .
\end{align}
However, we note that both frequencies exhibit negative growth rates ($1-\lambda -\nu \sim -\nu/Q$ and $-\lambda < 0$), preventing the expansion of either population. Thus, in this type of environmental selection, Adenines in the {\bf TR} are not counter-selected.

\subsection{$L>1$}
\begin{figure}[th!]
    \centering
    \includegraphics[width=0.8\columnwidth]{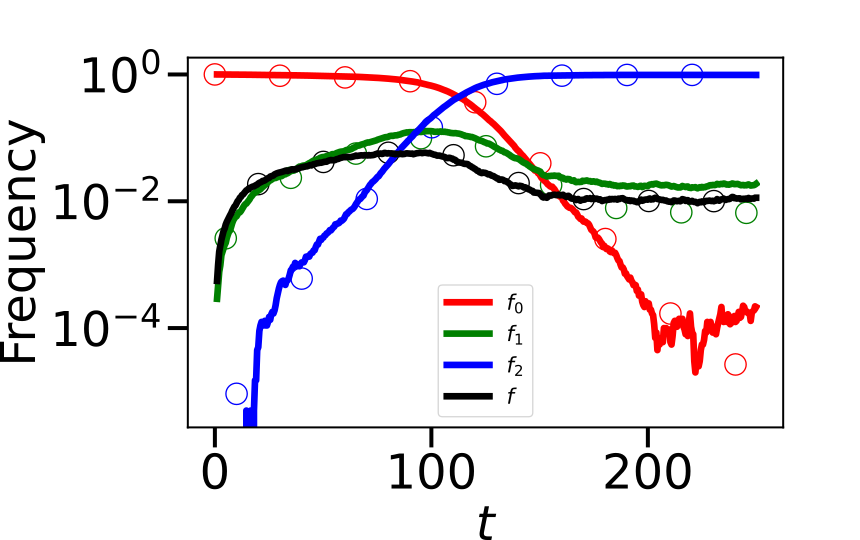}
    \caption{{\bf Frequency of each sub-population.} Comparison between numerical simulations (lines) and theoretical predictions (dots) of the fraction $f_k$ of the population of {\bf TR} having $k$ correct nucleotides and $L-k$ adenines, and $f(t)$ the rest. Parameters are $L=2$, $\mu=10^{-3}$, $\nu=10^{-1}$ and $N=4 \times 10^5$.}
    \label{fig:Pop_dyn}
\end{figure}

Fot this section, we build on the intuition developed for $L=1$: we consider that the time for the population with only $A$ at the {\bf TR} and with the optimal sequence at the {\bf VR} is much shorter than the time it will take to get mutations in the {\bf TR}. Then, we consider the dynamics of the {\bf TR} sequences which become progressively the same as the {\bf VR} sequences. Indeed, the cost of DGR is that sequences are replaced at rate $\nu$, thus sequences with the optimal sequence directly on their {\bf TR} have a fitness advantage of $\nu$.\\
We note $n_k(t)$ the number (resp. $f_k(t)$ the frequency) of the sequences with $k$ right bases on the {\bf TR} and $L-k$ Adenines. The rest of sequences is grouped together in a single population $n(t)$ that can not make the right {\bf VR} sequence under DGR. Initially, $n_0(t)=N$ (resp. $f_0(t)=1$), we only have $A$ in every {\bf TR}. We get the following system of equation which describes the {\bf TR} dynamics:
\begin{align}
\dot{n}_0(t)&=\left(L-\nu-\mu L\frac{Q-1}{Q} +\frac{\nu}{Q^{L}}\right)n_0+\frac{\mu}{Q}n_1 \\
    \dot{n}_k(t)&=\left(L-\nu-\mu L\frac{Q-1}{Q} +\frac{\nu}{Q^{L-k}}\right)n_k +\frac{\mu (L-k+1)}{Q}n_{k-1}+\frac{\mu (k+1)}{Q}n_{k+1} \\
    \dot{n}(t)&=(L-\nu)n(t)+\sum_{k=0}^L \frac{\mu L (Q-2)}{Q}n_k(t)\; .
\end{align}
We are interested in the regime where the population $n_0$ dominates, and define $\tau_c$ as the time where this hypothesis breaks. Thus, the equation for the frequencies become (using that the average fitness growth is mainly determined by the fitness of $n_0$):
\begin{align}
\dot{f}_0(t)&=\frac{\mu}{Q}f_1 \\
    \dot{f}_k(t)&=\frac{\nu}{Q^{L-k}}\left( 1-\frac{1}{Q^k}\right)f_k +\frac{\mu (L-k+1)}{Q}f_{k-1}+\frac{\mu (k+1)}{Q}f_{k+1} \\
    \dot{f}(t)&=\left(\mu L \frac{Q-1}{Q}-\frac{\nu}{Q^L}\right)f(t)+\sum_{k=0}^L \frac{\mu L (Q-2)}{Q}f_k(t) \; .
\end{align}
We check in \ref{fig:Pop_dyn} that this set of differential equations describe correctly the population dynamics simulation for the population size $N$ large enough (renormalizing the frequencies to 1 at every time step to account for the dynamics after $\tau_c$ as well).

Solving this system of equation, we get the time $\tau_c$ at which $\sum_{k=1}^Lf_k(t)+f(t)$ gets larger than $1/2$. \\
To understand the dependance of $\tau_c$ with the parameters, we consider two situations: either sequences with higher fitness (the ones corresponding to the $f_k(t)$'s) takeover, or the length is so large that fitness advantage of adding the right nucleotides is too low and random mutations in the {\bf TR} takeover (sequences corresponding to $f(t)$). \\ 
For the first regime, we neglect the transitions $f_{k+1} \to f_k$ which we suppose to be irrelevant to the determination of $\tau_c$. Besides,we consider $f_0(t)\approx 1$ fixed. This results in the simplified form for the triangular system of equations, 
\begin{align}
    \dot{f}_k(t)&=\frac{\nu}{Q^{L-k}}\left( 1-\frac{1}{Q^k}\right)f_k +\frac{\mu (L-k+1)}{Q}f_{k-1}  \; .
\end{align}
We can solve this sytem of equations in the following manner: for any $k$, $f_k(t)$ is a sum of exponentials with rates \begin{align}
    \lambda_j=\frac{\nu}{Q^{L-j}}\left(1-\frac{1}{Q^j}\right) \; ,
\end{align}
for $j=0$ to $j=k$, such that
\begin{align}
    f_k(t)=\sum_{j=0}^k C_{k,j} e^{\lambda_j t} \; .
\end{align}
In particular, using that for $k>1$,
\begin{align}
    f_k(t)=\frac{\mu(L-k+1)}{Q}\int_0^t f_{k-1}(s)e^{\lambda_k(t-s)} {\rm d}s \; ,
\end{align}
we deduce that for $j=0$ to $j=k-1$,
\begin{align}
    C_{k,j}=\frac{\mu(L-k+1)}{Q(\lambda_j-\lambda_k)} C_{k-1,j} \; ,
    C_{k,k}=\frac{\mu(L-k+1)}{Q}\sum_{j=0}^{k-1} \frac{C_{k-1,j}}{\lambda_k-\lambda_j}
\end{align}
which solution is given by
\begin{align}
    C_{k,j}=\frac{\prod_{m=1}^k \left(\frac{\mu(L-m+1)}{Q} \right) }{\prod_{m=1,m\neq j}^k (\lambda_j-\lambda_m) } \; .
\end{align}
In particular, when only considering the dominating exponential term at every $k$,
\begin{align}
    f_{k}(t)\approx \left(\frac{\mu L}{Q \lambda_k}  \right)^ke^{ \lambda_k t} \; .
\end{align}
Thus, $\tau_c$ is approximatively given by the smallest of the times $\tau_k$ such that $f_k(\tau_k)=1/2$, 
\begin{align}
    \tau_c \propto \min_{k=1,\ldots, L} \left\lbrace \frac{k Q^{L-k}}{\nu}\ln \left(\frac{\nu Q}{\mu k Q^{L-k}} \right) \right\rbrace \; .
\end{align}
Due to the exponential dependance in $L$, we can safely assume that it is the $k=L$ term which is the minimum, leading to
\begin{align}
    \tau_c \approx \frac{L}{\nu}\ln \left(\frac{\nu Q}{\mu L}\right). \label{eq:tau_c_1st}
\end{align}
In the second regime, we consider only $f(t)$, and the $f_0(t)\approx 1$ contribution (the others are of order $\mu$ at most), which results in 
\begin{align}
    f(t) &\approx \frac{\mu L \frac{Q-2}{Q}}{\mu L\frac{Q-1}{Q}-\frac{\nu}{Q^L}}\left[\exp \left[\left(\mu L\frac{Q-1}{Q}-\frac{\nu}{Q^L}\right) t\right]-1 \right] \\
    &\approx\frac{Q-2}{Q-2}\exp \left[\mu L\frac{Q-1}{Q} t\right]
\end{align}
leading to 
\begin{align}
    \tau_c \propto \frac{1}{\mu L} \label{eq:tau_c_2nd}
\end{align}
up to constant prefactors.\\
Combining Eqs.~\eqref{eq:tau_c_1st} and \eqref{eq:tau_c_2nd}, we obtain Eq.~(7) of the main text.

\newpage 

\bibliography{biblio}
\bibliographystyle{apsrev4-1}